\documentclass[aps, pre]{revtex4-1}

\usepackage{amssymb}
\usepackage{amsmath}
\usepackage{dsfont}
\usepackage{graphicx}
\usepackage{float}
\usepackage{appendix}

\usepackage{algorithm}
\usepackage[noend]{algpseudocode}

\usepackage{dblfloatfix}
\def\BState{\State\hskip-\ALG@thistlm}

\newcommand{\StatexIndent}[1][3]{%
  \setlength\@tempdima{\algorithmicindent}%
  \Statex\hskip\dimexpr#1\@tempdima\relax}
\algdef{S}[WHILE]{WhileNoDo}[1]{\algorithmicwhile\ #1}%

\begin{document}

\title{Revisiting entropy and entropy-related magnitudes of a symbolic sequence and their estimation}

\author{E. \surname{Arag\'on-P\'erez}}
\affiliation{Pleros, Parque Cient\'ifico Tecnol\'ogico de la Habana, Carretera a San Antonio, Boyeros. La Habana. Cuba.}

\author{A. \surname{Mesa-Rodr\'iguez}}
\affiliation{Universidad de La Habana, San Lazaro y L. CP 10400. La Habana. Cuba.}
\affiliation{MPI for the Physics of Complex Systems. N\"othnitzer Strasse 38. D-01187 Dresden.}
\altaffiliation{Pleros, Parque Cient\'ifico Tecnol\'ogico de la Habana, Carretera a San Antonio, Boyeros. La Habana. Cuba.}

\author{K. \surname{Garc\'ia-Medina}}
\affiliation{Universidad de La Habana, San Lazaro y L. CP 10400. La Habana. Cuba.}
\altaffiliation{Pleros, Parque Cient\'ifico Tecnol\'ogico de la Habana, Carretera a San Antonio, Boyeros. La Habana. Cuba.}

\author{D. \surname{Estevez-Moya}}
\affiliation{MPI for the Physics of Complex Systems. N\"othnitzer Strasse 38. D-01187 Dresden.}
\altaffiliation{Pleros, Parque Cient\'ifico Tecnol\'ogico de la Habana, Carretera a San Antonio, Boyeros. La Habana. Cuba.}

\author{A. \surname{Mollineda-Rodríguez}}
\affiliation{Universidad de La Habana, San Lazaro y L. CP 10400. La Habana. Cuba.}
\affiliation{Pleros, Parque Cient\'ifico Tecnol\'ogico de la Habana, Carretera a San Antonio, Boyeros. La Habana. Cuba.}

\author{E. \surname{Estevez-Rams}}
\email{estevez@fisica.uh.cu}
\affiliation{Universidad de La Habana, San Lazaro y L. CP 10400. La Habana. Cuba.}
\affiliation{Pleros, Parque Cient\'ifico Tecnol\'ogico de la Habana, Carretera a San Antonio, Boyeros. La Habana. Cuba.}

\begin{abstract}
This article revisits entropy and related magnitudes—entropy density, effective measure complexity, and informational distance—for the analysis of sequential data. We present their mathematical foundations within information theory and ergodic theory, establishing connections between probabilistic Shannon entropy and algorithmic Kolmogorov complexity. A central result, Ziv’s theorem, demonstrates the asymptotic equivalence between the normalised number of distinct words and entropy density, providing the theoretical basis for estimation via Lempel-Ziv factorisation. We examine the convergence properties of these magnitudes and introduce estimation procedures for finite sequences, including a method for the effective measure complexity and the transfer entropy rate. An open-source software library implementing parallel algorithms for efficient Lempel-Ziv factorisation is presented. The practical utility of the framework is illustrated through three case studies: functional brain connectivity analysis from fMRI data, characterisation of spatiotemporal dynamics in coupled nonlinear oscillators, and electrocardiogram-based classification of atrial fibrillation, demonstrating how entropic measures capture meaningful structural and dynamical properties across diverse domains.
\end{abstract}

\date{\today}

\maketitle

\begin{quotation}
Many complex systems are known to us through a single recorded signal, from which regularity, memory, and randomness must be disentangled. Entropy density, effective measure complexity, and informational distance measure exactly that -- how unpredictable a signal is, how much memory it carries, how far two signals lie apart -- yet their theory sits scattered across decades of literature, and their estimation is frequently done wrong. This article assembles that theory in one place, links Shannon's entropy to Kolmogorov's algorithmic complexity, and shows how Lempel-Ziv factorisation turns both into practical, parameter-free estimators, backed by open-source software that processes sequences of a billion symbols. Applied unchanged to fMRI recordings, a ring of coupled nonlinear oscillators, and electrocardiograms, the same measures recover task-related brain activity, detect collective behaviour near the onset of chaos, and separate healthy from arrhythmic heart rhythms with clinically relevant accuracy, showing that a single information-theoretic toolkit captures memory and unpredictability across strikingly different systems.
\end{quotation}

\section{Introduction}

Data from complex dynamical systems often appear as one or a few time series, from which the system's dynamics must be extracted \cite{kantz97}. In data analysis, the usual aims are to assess regularity versus randomness, to identify patterns, to discover governing laws, or to predict future system behaviour \cite{crutchfield12}. One useful approach is to treat the dynamical system as an information source and apply information-theoretic tools alongside statistical descriptions \cite{coverElementsInformationTheory2006}. In general, a single or just a few data sets are insufficient to fully understand the underlying dynamical system that drives the information source, yet these are often the only available data. In such cases, assuming ergodicity becomes necessary to extract the relevant information from the data \cite{grassbergerQuantitativeTheorySelfgenerated1986}.

For most systems, a fundamental quest is to find the relation between randomness and pattern in the system signal \cite{crutchfieldRegularitiesUnseenRandomness2001}. Randomness can be considered noise or, in a more useful perspective, the intrinsic uncertainty of the system, while a pattern is related to the deterministic laws that govern the system's behaviour. It has been argued that for a complex dynamical system, the behaviour lies between complete predictability and total randomness \cite{crutchfieldRegularitiesUnseenRandomness2001}. Entropy density captures the irreducible unpredictability in a signal. Entropy density, also known as Kolmogorov-Sinai entropy, entropy rate, among other names, has been defined in several fields, including information theory, ergodic theory, and dynamical systems \cite{coverElementsInformationTheory2006,badii99}.  A lesser-used measure of complexity is the effective measure of complexity, introduced by Grassberger \cite{grassbergerQuantitativeTheorySelfgenerated1986}, which aims to capture a system's intrinsic memory and, in time-series analysis, its reliance on the past to define its future. Both magnitudes can be theoretically justified by Kolmogorov's algorithmic complexity, which laid a solid foundation for the analysis of data beyond the ensemble-statistical point of view \cite{kolmogorov65}. Kolmogorov's theory of complexity builds on the existence of a universal computational machine, as demonstrated by Turing, and on the computational theory of recurrent functions \cite{vitanyi93}. From Kolmogorov theory, several concepts emerge that are closely related to entropy density and effective measure complexity; significantly, it also allows us to define the informational distance as a measure of the distance between two data sets \cite{liSimilarityMetric2004}.

The informational theory related concepts have been used extensively in analyzing experimental and theoretical models and there is an extensive literature regarding their use  \cite{badii99,d.gusevComplexityMeasuresGenetic1999,ferrarioComplexityAnalysis242004,coverElementsInformationTheory2006,abasoloNonlinearAnalysisIntracranial2007,feldman08,chelaniComplexityAnalysisCO2011,crutchfield12}. In particular there is abundant literature dealing with the estimation of entropy density, and in a much lesser extent, effective measure complexity and informational distance \cite{kaspar87,gilbertLempelZivAlgorithmMessage1992,baruchFilteringPredictionIndividual2000,alfonsecaCommonPitfallsUsing2005,cebrinNormalizedCompressionDistance2007,lesne09,melchertAnalysisPhaseTransition2015,avineryUniversalAccessibleEntropy2019}. In spite of that, there persist some confusion in the validity of the estimation procedures, specially in the latter magnitudes, and its use, and the mathematics of the whole subject appears scattered in the scientific literature, some misconceptions also persist \cite{lesne09,grassbergerCommentsComputationalMechanics2018}.

This article revisits key concepts and their mathematical foundations, including estimation and practical use. It gathers work previously scattered across various sources, including new, unpublished results from the authors. The aim is to offer a comprehensive treatment, from theory to practice, in a single source. Publicly available software and a library for estimating these magnitudes are also introduced.

What this article is not, is a review, there are some important areas left out, specially the early work of Grassberger \cite{grassbergerQuantitativeTheorySelfgenerated1986,grassbergerCommentsComputationalMechanics2018} and the further studies by Crutchfield and Feldman \cite{crutchfieldRegularitiesUnseenRandomness2001} on entropy estimation for finite state automata, although some analytical results from their work will be used to compare estimated values of entropy density, and effective measure complexity with the exact values.

The article comprises three main parts: (1) mathematical foundations of information theory (sections \ref{sec:notation}, \ref{sec:ent}, \ref{sec:emc}, \ref{sec:dist}); (2) analysis of Lempel-Ziv factorisation, complexity, and entropic magnitude estimation (sections \ref{sec:lz}, \ref{sec:est}, \ref{sec:algol}); and (3) case studies (section \ref{sec:case}). Conclusions follow. The appendix contains additional material, including proofs and explanations of procedures. Table~\ref{tab:notation} collects the notation used throughout.

\section{Strings, ergodicity and stationarity\label{sec:notation}}
The intuitive notion of a string corresponds to a sequence of successive symbols, which take values from a finite set, such as a binary alphabet. The index that traverses the positions of the symbols in a sequence is called time. Although in practice one must deal with finite strings, the concept extends to infinite strings, and this extension is used in proving certain theorems. However, it is worth clarifying that there are several types of infinite strings; the most general ones never begin or end (e.g., $\ldots 1011010000111010100110011111110\ldots$), and are called bi-infinite. If a point is taken in a bi-infinite string, to the right there will be an infinite string that has a starting point but never ends (e.g., $abbbaccbaab\ldots$); and to the left we will have one that ends but does not have a starting point ($\ldots32214441233$). Any point in a bi-infinite string can be taken as the middle point.

Formally, a string is a successive concatenation of symbols $u=\ldots u_{-2}u_{-1} u_0 u_1u_2\ldots u_i\ldots$ emitted by a symbol source. $u_i$ is known as a character and takes values from an alphabet $\mathcal{A}$, $u_i\in\mathcal{A}; \forall i\in\mathbb{Z}$. The index $i$ will be called time even though it does not represent that physical variable. The number of symbols available to the source will be taken as finite, $|\mathcal{A}|<\infty$; the cardinality of the alphabet will be denoted by $\alpha(=|\mathcal{A}|)$. Throughout, string positions are indexed from zero; $u^{(\ell)}=u_0u_1\ldots u_{\ell-1}$ denotes the prefix of length $\ell$. Unless otherwise stated, $\log\equiv\log_\alpha$; if the sequence is binary, then $\alpha=2$ and entropies are expressed in bits.

When a string is finite, we call it a word. A word $w$ has length $|w|$. It is important to include the null sequence $\Lambda$, which has length $(|\Lambda|=)0$. If we have two strings $w$ and $u$, the sequence $wu$ is the concatenation of both sequences in the indicated order, and its length would be $|w|+|u|$.

It is convenient to consider a symbol source or emitter as generating discrete strings uninterruptedly; its output is bi-infinite strings. To circumvent the problems associated with time in generating bi-infinite strings when considering the occurrence of multiple strings, we can imagine that instead of a single emitter, we have a set, called an ensemble, of countably infinite identical emitters, each generating a bi-infinite string. Since, from a practical point of view, it is useful to consider finite strings, they are taken as the result of observing an emitter’s output over a finite time interval ($\ldots100100000\boxed{11000111111}10010000111\ldots$). For a source, the set of rules it follows as they generate the bi-infinite string is called a grammar and denoted by $G$ (e.g., a binary source that only emits an even number of 0s).

The set of all strings of finite length (words) will be denoted by $\mathcal{A}^*$ and those of infinite length by $\mathcal{A}^\infty$.

Consider an infinite string $u\in\mathcal{A}^{\infty}$. If we define the set of all finite substrings of $u$, it is possible to define a $\sigma$-algebra on that set and assign a probability measure to the set. 

For example, take the set of all strings of finite length. Consider the initial collection formed by subsets of strings, where each subset contains strings of the same length. From this initial collection, a $\sigma$-algebra can be generated, and probabilities can be assigned to the strings based on their length. A similar probability assignment can be done to other partitions of the strings into classes.

A subsequence or substring $w_a^b\subseteq u$ is a sequence of symbols extracted from $u$ starting at time $a$, up to time $b$, and, therefore, of length $b-a+1$. The word $w$, where the position in the $u$ string is irrelevant, is called the base word. A subsequence can be infinite or can be a word.

It is useful to introduce the drop operator $\pi$, such that when operating on the left on a word, it removes the first character of the string, $\pi w_a^b\equiv w_{a+1}^b$ (e.g., $\pi 101100=01100$); and operating on the right removes the last character of the string, $w_a^b \pi\equiv w_a^{b-1}$ (e.g., $101100\pi=10110$).

A sequence can be shifted to the right such that if we call $\Theta[u]$ the shifted string, we will have $[\Theta(u)]_t = u_{t-1}$ ($\Theta[\dots u_5 1 0 0 1 u_{10} \ldots] = \dots u_4 u_5 1 0 0 1 \ldots$). We define the same for a left shift ($\Theta^{-1}[u]$) where we will have $[\Theta^{-1}(u)]_t = u_{t+1}$ ($\Theta^{-1}[\dots u_5 1 0 0 1 u_{10} \ldots] = \dots 1 0 0 1 u_{10} u_{11} \ldots$). The operator $\Theta$ is called the right shift operator, and $\Theta^{-1}$ is called the left shift operator.

The two most important properties required of a source are ergodicity and stationarity. 

Ergodicity means that the strings emitted by the source are statistically indistinguishable: for almost every string $u\in\mathcal{A}^\infty$ emitted by an ergodic source $\psi$, the relative frequency with which any finite word $w$ occurs along $u$ converges to the same value $P(w)$, whatever the string (e.g., $P_u(w)=P_v(w)$ for two emitted strings $u$ and $v$). A single string then characterises the source, and ensemble averages can be replaced by averages along one realisation.

Stationarity refers to the temporal invariance of the emitter's statistical properties. In other words, the probability of a specific finite string, of any length, occurring in a bi-infinite string does not depend on the instant in time at which the finite string occurs.

Stationarity implies that the probability of occurrence of a substring in an infinite string emitted by the source depends only on the base word of the substring and not on the starting time. A stationary source implies that for all $u\in\mathcal{A}^\infty$, $P(u)=P(\Theta(u))$.

If a process is stationary, all instances $w_i^j$ of the base word $w$ have the same probability ($P(w_a^b)=P(w)$) for any string $u\in\mathcal{A}^\infty$.

\begin{table*}[!t]
\caption{\label{tab:notation}Notation used throughout. A hat denotes a magnitude estimated from a finite sequence.}
\begin{ruledtabular}
\footnotesize
\renewcommand{\arraystretch}{1.08}
\begin{tabular}{l@{\hspace{0.6em}}p{0.315\textwidth}@{\hspace{1.2em}}l@{\hspace{0.6em}}p{0.315\textwidth}}
 
\multicolumn{2}{l}{\textit{Alphabet, strings and words}} &
\multicolumn{2}{l}{\textit{Algorithmic complexity and distance}}\\[2pt]
 
$\mathcal{A},\ \alpha$      & Alphabet and its size, $\alpha=|\mathcal{A}|$ &
$\mathcal{U}$               & Universal Turing machine \\
 
$\mathcal{A}^{*},\ \mathcal{A}^{\infty}$ & Words; infinite strings &
$K(w),\ w^{*}$              & Kolmogorov complexity; minimal program \\
 
$\psi,\ G$                  & Symbol source; its grammar &
$K(w_1|w_2),\ K(w_1{:}w_2)$ & Conditional complexity; algorithmic mutual information \\
 
$u,\ v$                     & Bi-infinite string emitted by $\psi$ &
$k_\ell,\ k(u)$                & Complexity per symbol, $K(u^{(\ell)})/\ell$ \\
 
$u_i$                       & Character at time $i$, $u_i\in\mathcal{A}$ &
$\eta_\ell$                    & Any of $\zeta_\ell,h_\ell,\Delta H_\ell,k_\ell$ \\
 
$u^{(\ell)}$                   & Prefix $u_0\ldots u_{\ell-1}$, of length $\ell$ &
$\delta h^{(\eta)}_\ell,\ E_\eta$ & Deficit $|\eta_\ell-h|$; its sum over $\ell$ \\
 
$w,\ |w|$                   & Word (finite string) and its length &
$\nu,\ \beta$               & Decay exponents of $\delta h^{(\eta)}_\ell$ \\
 
$w^{b}_{a}$                 & Substring of $u$ from time $a$ to $b$ &
$d_{\parallel},\ d$         & Absolute and normalised information distance \\
 
$\Lambda$                   & Null string, $|\Lambda|=0$ &
$\Delta K,\ \Delta h$       & Net flow of information \\
 
$u^{-},\ u^{+}$             & Past and future halves of $u$; $u^{\mp}_\ell$ the $\ell$ symbols nearest the cut &
$TE,\ TER$                  & Transfer entropy and its per-symbol rate \\
 
$N,\ \ell$                     & Sequence length; block length &
$m,\ M$                     & Embedding order; state-vector dimension, $M=2m+1$ \\
 
$p$                         & Period of a periodic sequence &
\multicolumn{2}{l}{}\\[4pt]
 
\multicolumn{2}{l}{\textit{Operators and probability}} &
\multicolumn{2}{l}{\textit{Lempel--Ziv factorisation and estimation}}\\[2pt]
 
$\pi$                       & Drop operator, $\pi w^{b}_{a}=w^{b}_{a+1}$, $w^{b}_{a}\pi=w^{b-1}_{a}$ &
$\mathcal{V}(w)$            & Vocabulary: all distinct words of $w$ \\
 
$\Theta,\ \Theta^{-1}$      & Right and left shift operators &
$F_i,\ a_i$                 & $i$-th factor; position where it begins \\
 
$P[\cdot]$                  & Probability of a string &
$LZ(w),\ C_{LZ}$            & Exhaustive factorisation; its number of factors \\
 
$\mathbb{E}[\cdot]$         & Expected value over the source &
$c_{LZ}(N)$                 & Normalised complexity, $C_{LZ}/(N/\log N)$ \\
 
$\mathcal{N}(\mu,\sigma^2)$ & Normal distribution &
$n_F$                      & Number of factors of a factorisation \\
 
$\Omega$                    & Typical set of the AEP &
$\lambda_k,\ \mu,\ \sigma$  & Factor length; its mean and standard deviation \\
 
$\epsilon$                  & Arbitrarily small positive quantity &
$Z$                         & Variate drawn from $\mathcal{N}(0,1)$ \\
 
$\log$                      & $\equiv\log_\alpha$ unless otherwise stated &
$\epsilon_N,\ \kappa$       & Finite-size correction; offset of the $C_{LZ}$ ansatz \\[4pt]
 
\multicolumn{2}{l}{\textit{Entropies}} &
$n,\ N_{\mathrm{eff}}$      & Truncation limit of $\widehat E(n)$; effective length $Nh$ \\[2pt]
 
$H_\ell$                       & Shannon block entropy of the length-$\ell$ words &
$m_\ell,\ n_i,\ f_{1,\ell}$       & Windows sampled; block counts; blocks seen once \\
 
$\Delta H_\ell$                & Entropy gain, $H_\ell-H_{\ell-1}$ &
$\widehat C_\ell,\ \widehat p_i,\ \widehat H^{CS}_\ell$ & Coverage; adjusted probability; Chao--Shen entropy \\
 
$h$                         & Entropy density (Kolmogorov--Sinai entropy) &
$\Gamma$               & Number of surrogates \\
 
$h_\ell$                       & Finite entropy rate of order $\ell$, $H_\ell/\ell$ &
$v_t,\ z_t,\ \tilde\alpha$  & State vector; its encoding; extended alphabet size $\alpha^{M}$ \\[4pt]
 
$\#_\ell$                      & Number of distinct words of length $\ell$ &
\multicolumn{2}{l}{\textit{Model sources and case studies}}\\[2pt]
 
$\zeta_\ell$                   & Defined by $\#_\ell=\alpha^{\,\ell\zeta_\ell}$ &
$S_i,\ \mathcal{T}^{a}_{i}$ & FSA state; its transition probability \\
 
$I[\cdot,\cdot]$            & Mutual information &
$p^{(t)}_i,\ p^{(\infty)}_i$ & Probability of state $S_i$ at step $t$; asymptotic value \\
 
$E$                         & Effective measure complexity, $E=I[u^{-},u^{+}]$ &
$\tau,\ \bar\tau$           & Transition probability of the even and 2ssp processes \\
 
$\ell_c$                       & Length beyond which $\Delta H_\ell$ has converged to $h$ &
$p,\ \bar p$                & Symbol probability of the RRXOR process \\
 
\multicolumn{2}{l}{} &
$\theta_i,\ \omega,\ \gamma$ & Oscillator phase, natural frequency, coupling \\
 
\multicolumn{2}{l}{} &
$r_{ij},\ \oplus$           & Recurrence-matrix entry; exclusive-or \\
 
\end{tabular}
\end{ruledtabular}
\end{table*}

\section{Block entropy, entropy gain and entropy density\label{sec:ent}}

\subsection{Shannon block entropy}

Consider a bi-infinite string $u$ from which the set $\{w^{(\ell)}\}$ of all finite different substrings, $w^{(\ell)}$, of length $\ell$ are taken. Let $\#_\ell$ be the cardinality of  $\{w^{(\ell)}\}$. To each string $w^{(\ell)}$ a probability $P(w^{(\ell)})$ is assigned. The Shannon block entropy is then defined as \cite{shannon48,shannon51,coverElementsInformationTheory2006}
\begin{equation}
H_\ell(u)=-\sum\limits_{w^{(\ell)}\in u}P(w^{(\ell)})\log P(w^{(\ell)}). \label{eq:block}
\end{equation}
The summand $0\log 0\equiv 0$ by definition. $H_\ell(u)=0$ if $u$ is a constant string formed by a single character and $H_\ell(u)$ attains its maximum value $\log \#_\ell$  when all different words $w^{(\ell)}$ have the same probability $P(w^{(\ell)})=1/\#_\ell$.

$H_\ell(u)$ is a non- decreasing function of $\ell$: $H_{\ell+1}(u)\ge H_{\ell}(u)$ (see Appendix \ref{app:bentropy}).

\subsection{Entropy gain}

The entropy gain, a non-negative magnitude, is defined as \cite{crutchfieldRegularitiesUnseenRandomness2001}
\begin{equation}
\Delta H_\ell(u)\equiv H_\ell(u)-H_{\ell-1}(u).\label{eq:entgain}
\end{equation}
Since the null word $\Lambda$ occurs with probability one, $H_0(u)=0$ and $\Delta H_1(u)=H_1(u)$. Equation (\ref{eq:entgain}) does not define $\Delta H_0$; following Crutchfield and Feldman \cite{crutchfieldRegularitiesUnseenRandomness2001}, we adopt the convention $\Delta H_0(u)=\log\alpha=1$, the uncertainty assigned to a symbol before any observation is made. From equation (\ref{eq:hhh})
\begin{equation}
 \Delta H_\ell(u)=-\sum\limits_{w^{(\ell)}\in u} P(w_0\ldots w_{\ell-1}) \log P(w_{\ell-1} | w_0,...,w_{\ell-2})\equiv H(u_{\ell-1}|u^{(\ell-1)}). \label{eq:entgain1}
\end{equation}
Equations (\ref{eq:entgain}) and (\ref{eq:entgain1}) allow two equivalent interpretations of the entropy gain. On one hand, it is an estimate of how random a symbol source looks when only substrings up to length $\ell$ are taken into account. On the other hand, it is the mean uncertainty of the symbol $u_{\ell-1}$ if the prefix $u^{\ell-1}$ is known. In the last sense, it is the amount of information that has to be stored in order to make an optimal prediction of $u_{\ell-1}$ \cite{crutchfieldRegularitiesUnseenRandomness2001}.

For a stationary process
\[
\Delta H_\ell(u)=H(u_{\ell-1}|u^{(\ell-1)})\leq H(u_{\ell-1}|\pi u^{(\ell-1)})=H(u_{\ell-2}|u^{(\ell-2)})=\Delta H_{\ell-1}(u).
\]
The inequality follows from the fact that conditioning reduces the entropy, and the equality results from the stationarity condition. $\Delta H_\ell(u)$ is a non-increasing function of $\ell$.

As $\Delta H_\ell(u)\geq 0$, the last property implies
\begin{equation}
0 \leq \Delta H_\infty(u)<\infty,\label{eq:0dh}
\end{equation}
where
\begin{equation}
\Delta H_\infty(u)=\lim\limits_{\ell\rightarrow \infty}\Delta H_\ell(u)=\lim\limits_{\ell\rightarrow \infty}  H\left[u_\ell|u^{(\ell)}\right].\label{eq:hdi}
\end{equation}
The limit $\Delta H_\infty(u)$ exists for an ergodic stationary source. In what follows, when needed, the dependence on $u$ will be dropped from $\Delta H_\ell$, since, for an ergodic source, any infinite string $u$ is statistically equivalent to any other.

For a periodic sequence of minimal period $p$, the probabilities are the frequencies of occurrence along the sequence, and the words of length $\ell\geq p$ are the $p$ cyclic phases of the period, each occurring exactly once per period; their distribution is therefore uniform, and $H_\ell=\log p$ for $\ell\geq p$, from where
\begin{equation}
\Delta H_\ell=\Delta H_\infty=0. \;\;\;(\ell\geq p), \label{eq:deltahp}
\end{equation}
for a random sequence, by which we mean throughout a realisation of an i.i.d.\ source with equiprobable symbols, $P(a)=1/\alpha$ for all $a\in\mathcal{A}$, $H_\ell=\ell$ and
\begin{equation}
\Delta H_\ell=\Delta H_\infty=1.\label{eq:deltahr}
\end{equation}

\subsection{Entropy density}

To measure the mean uncertainty per symbol of the ergodic source, the entropy density is defined as \cite{shannon48,shannon51}
\begin{equation}
h\equiv \lim\limits_{\ell\rightarrow \infty}\frac{H_\ell}{\ell},\label{eq:h}
\end{equation}
Entropy density has been referred to by different names in different contexts, such as Kolmogorov-Sinai (KS) entropy or entropy rate \cite{badii99,coverElementsInformationTheory2006}.

Using the chain rule
\begin{equation}
\frac{H_\ell(u)}{\ell}=\frac{1}{\ell}\sum\limits_{i=0}^{\ell-1}H(u_i|u^{(i)}),\label{eq:cadena}
\end{equation}
and using the Ces\'aro theorem, which states that if there is a sequence in $n$ such that $a_n\rightarrow a$, and $b_n=1/n\sum_{i=0}^{n-1}a_i$, then $b_n\rightarrow a$, we have \cite{crutchfieldRegularitiesUnseenRandomness2001},
\begin{equation}
h=\lim\limits_{\ell\rightarrow \infty}\frac{H_\ell}{\ell}=\lim\limits_{\ell \rightarrow \infty}H(u_\ell|u^{(\ell)})=\Delta H_\infty.\label{eq:hdeltah}
\end{equation}
This last expression allows us to understand the entropy density as the uncertainty in a new symbol when an infinite number of symbols in the string have been measured. In expression (\ref{eq:h}), the numerator is a non-decreasing function of $\ell$, while the denominator is strictly increasing; knowing that for an ergodic and stationary source, the limit exists, convergence is guaranteed to be uniform, and
\begin{equation}
H_\ell\sim h \ell\;\;\;\ell\rightarrow \infty.
\end{equation}
For a large enough $\ell$, $H_\ell(u)$ is a linear function of $\ell$ with slope equal to the entropy density \cite{crutchfieldRegularitiesUnseenRandomness2001}.

From equation (\ref{eq:h}), we can define the terms in the sequence to infinity as
\begin{equation}
h_\ell=\frac{H_\ell}{\ell}, \label{eq:hl}
\end{equation}
sometimes known as the finite entropy rate of order $\ell$. Although equation (\ref{eq:hdeltah}) guarantees that in the limit $\ell\rightarrow \infty$, $\Delta H_\ell$ and $h_\ell$ are equal, for finite $\ell$ this is not the case. In fact, the entropy gain converges faster than $h_\ell$ (See Appendix \ref{app:convergence}).

For a periodic sequence of minimal period $p$, $H_\ell=\log p$ for $\ell\geq p$, from where
\begin{equation}
h_\ell=\frac{\log p}{\ell} \;\;\;(\ell\geq p), \label{eq:hlp}
\end{equation}
and $h=0$.
For a random sequence
\[
h_\ell=h=1, 
\]
the same value as $\Delta H_\ell$.

\subsection{Number of different words\label{sec:nopd}}

As before, let $\#_\ell=\alpha^{\ell \zeta_\ell(u)}$ be the number of different words of length $\ell$ in a string $u$.

If for a given $\ell$ we have $\#_\ell$, then at most we will have $\alpha \#_\ell$ different words of length $\ell+1$. It follows
\[
\alpha^{\ell \zeta_\ell(u)}\alpha\ge \alpha^{(\ell+1) \zeta_{\ell+1}(u)},
\]
from where
\begin{equation}
\zeta_{\ell+1}\leq\frac{\ell}{\ell+1} \zeta_\ell(u)+\frac{1}{\ell+1}.\label{eq:hl1}
\end{equation}
On the other hand, $\#_{\ell+1}\ge \#_\ell$ and
\begin{equation}
\zeta_{\ell+1}\ge \frac{\ell}{\ell+1} \zeta_\ell(u).\label{eq:hl1b}
\end{equation}
There are at most $\alpha^\ell$ different words of length $\ell$, therefore
\begin{equation}
\zeta_\ell(u)\le 1.\label{eq:hsigma}
\end{equation}
By definition $\#_\ell(u)\ge 1$, which implies that
\begin{equation}
\zeta_\ell(u)\ge 0. \label{eq:hcero}
\end{equation}
Form inequalities (\ref{eq:hsigma}) and (\ref{eq:hcero}) it follows that $ \zeta_\ell(u)$ is bounded from below and above
\begin{equation}
0\le  \zeta_\ell(u) \leq 1. \label{eq:hpapa}
\end{equation}
Consequently the $\liminf\limits_{\ell\rightarrow \infty} \zeta_\ell(u)$ and $\limsup\limits_{\ell\rightarrow \infty} \zeta_\ell(u)$ exist, are finite and are equal (see Appendix \ref{app:limsup}).

The relation
\begin{equation}
\zeta_\infty(u)=\lim\limits_{\ell\rightarrow \infty}  \zeta_\ell(u)\leq 1,\label{eq:hhl}
\end{equation}
makes $\zeta_\infty(u)$ a quantity that characterises an infinite sequence $u$, and thus, for an ergodic source, the source itself. As done before, the dependence of $\zeta_\ell$ on $u$ will be dropped when it clutters the notation.

For a periodic sequence of minimal period $p$ the different words of length $\ell\geq p$ are the $p$ cyclic phases of the period, so $\#_\ell= p$ for $\ell\geq p$, from where it follows that
\begin{equation}
\zeta_\ell= \frac{\log p}{\ell}\;\;\;(\ell\geq p),\label{eq:zetap}
\end{equation}
and consequently $\zeta_\infty=0$.
For a random sequence, and more generally for any i.i.d.\ sequence in which every symbol has non-zero probability, $\#_\ell=\alpha^\ell$, and
\begin{equation}
\zeta_\ell=1. \label{eq:zetar}
\end{equation}

\subsection{Number of distinct words and entropy}

Consider an ergodic source characterized by an entropy rate $h$, let $u$ be an infinite string produced by the source,  Ziv \cite{ziv78} proved (see Appendix \ref{app:ziv78})
\begin{equation}
P\left[\zeta_\infty=h\right]=1.\;\;\;\;\; \text{(Ziv’s theorem 5)}\label{eq:ziv5}
\end{equation}

In an ergodic source, its statistical properties can be estimated accurately from a single very long sequence generated by the source. The question is, how do we extract these properties from a single sequence? Equation (\ref{eq:ziv5}) tells us one way. $\zeta_\infty$ is, in principle, computable from a sufficiently long sequence. But beyond that practical value, equation (\ref{eq:ziv5}) points to something more fundamental: asymptotically, the number of different strings of length $\ell(\gg 1)$ is determined by the entropy of the ergodic source. In retrospect, it should not surprise us; a source is more unpredictable the more distinct strings it can generate. But the result also surprises us; entropy, and correspondingly $h$, is a function of occurrence probabilities, but $\zeta_\infty$ is not a function of how many times each different string appears; it only depends on the number of distinct strings.

Ziv's theorem (\ref{eq:ziv5}) guarantees the equivalence of $\zeta_\ell$ to $h_\ell$ and $\Delta H_\ell$ at the limit $\ell\rightarrow \infty$; it does not say anything for finite values of $\ell$. From equation (\ref{eq:block}) and the fact that the block entropy reaches its maximum value for a uniform distribution $P(u)=1/\#_\ell(u)$, it follows.
\[
\log \#_\ell=\ell \zeta_\ell(u)\geq H_\ell(u),
\]
that is, $\zeta_\ell\geq h_\ell$. Together with $h_\ell\geq \Delta H_\ell$, Eq.~(\ref{eq:faster}) of Appendix~\ref{app:convergence}, and $\Delta H_\ell\geq 0$, it follows that
\begin{equation}
\zeta_\ell \geq h_\ell \geq \Delta H_\ell \geq 0. \label{eq:hlzetal}
\end{equation}

\subsection{Algorithmic complexity and entropy}

Kolmogorov complexity, also known as algorithmic complexity, quantifies the information content of a finite string \cite{vitanyi93}. Formally, for a given Universal Turing Machine (UTM) $\mathcal{U}$ and a string $w$, this measure corresponds to the length of the shortest possible program $w^*$ that can generate $w$ as output:
\begin{equation}
K_{\mathcal{U}}(w) = \min_{w^* \in \mathcal{A}^*} \bigl\{ |w^*| \mid \mathcal{U}(w^*) = w \bigr\},\label{eq:kolmogorov}
\end{equation}
where $K_{\mathcal{U}}(w)$ is the Kolmogorov complexity,  $|w^*|$ denotes the length of program $w^*$,  $\mathcal{U}(w)$ represents the output produced when $\mathcal{U}$ executes program $w^*$. As changing the UTM amounts to adding a constant to the Kolmogorov complexity regardless of the string $w$, the use of another UTM just adds an $O(1)$ upper-bound correction to the defined magnitudes; the subscript $\mathcal{U}$ can be dropped, and in what follows we use $K(w)$. Kolmogorov complexity of a string represents its inherent (irreducible) information content.

There are certain constraints needed for the shortest possible program $w^*$. First of all, it must halt after a finite number of steps, and it will also be required that once it halts, its only output must be the string $w$. Another convenient condition usually imposed on $w^*$ is that it must be self-delimiting, that is, $w^*$ carries with it the information about its size, which implies an additional correction of $\log |w|$, we will drop the requirement in what follows for clarity. 

The joint algorithmic complexity of strings $w_1$ and $w_2$ is defined as the size of the smallest program that computes both strings. The following inequality holds,
\begin{equation}
K(w_1,w_2)\leq K(w_1)+K(w_2)+O(1).\label{eq:jointk}
\end{equation}
The conditional Kolmogorov complexity $K(w_1|w_2)$ is understood as the length of the minimal self-delimiting program that computes the string $w_1$ if the additional string $w_2$ is provided as input. The following relation holds,
\begin{equation}
K(w_1,w_2)=K(w_2)+K(w_1|w_2)+O(1).\label{eq:condk}
\end{equation}
Finally, the algorithmic mutual information can be taken as
\begin{equation}
K(w_1:w_2)=K(w_1)+K(w_2)-K(w_1,w_2),\label{eq:mutualk}
\end{equation}
which measures the additional amount of information needed to  calculate $w_1$ and $w_2$ separately instead of jointly. Inequality ($\ref{eq:jointk}$) makes $K(w_1:w_2)\geq 0$.

To extend the definition of $K(\cdot)$ to infinite string $u$, consider the finite string $u^{(\ell)}$ of length $\ell$, the algorithmic complexity per symbol can then be defined as
\begin{equation}
k(u)=\lim\limits_{\ell\rightarrow \infty} k_\ell=\lim\limits_{\ell\rightarrow \infty} \frac{K(u^{(\ell)})}{\ell}.\label{eq:k}
\end{equation}
There are several troublesome subtleties in the above definition, particularly the question of whether such a limit exists. A more sound definition would be to define two $k(u)$, one taking the $\limsup$, $k_{sup}(u)$, and another taking the $\liminf$, $k_{inf}(u)$, but as further on, we will be dealing exclusively with finite strings, we will leave it as defined by (\ref{eq:k}) and the reader is referred to appropriate reference for the finer details \cite{vitanyi93}.

While Shannon entropy quantifies information through the probabilistic characteristics of data sources, treating individual objects as statistical realisations, Kolmogorov complexity evaluates information content directly through an object’s algorithmic properties, measuring its inherent descriptive complexity. This conceptual divergence notwithstanding, a systematic correspondence emerges between Shannon-theoretic and Kolmogorov-theoretic frameworks: expectations of algorithmic information measures typically match classical entropy quantities, establishing fundamental equivalences between the two approaches to information quantification. This fact, for an ergodic source, should come as no surprise, as the statistical properties of the source are translated to those of a single string.

The main result, proven in  \cite{vitanyi93}, is that given an ergodic source and a set ${u}$ of strings from the source with a probability measure defined over the set, then
\begin{equation}
\lim\limits_{\ell\rightarrow \infty}\mathbb{E}\left [\frac{K(u^{(\ell)})}{\ell}\right]=\lim\limits_{\ell\rightarrow \infty}\mathbb{E}\left [k_\ell\right]=h,
\end{equation}
where $\mathbb{E}[\cdot]$ is the expected value. The main shortcoming of Kolmogorov complexity lies in the fact that it is non-computable due to the unsolvable nature of the halting problem, which renders it impossible to find a recursive function that can determine the shortest algorithm that produces a given string  \cite{vitanyi93}. Yet the non-computability of $K(w)$ does not necessarily translate to non-computability of $K(u^{(\ell)})/\ell$ \cite{zurek89a}.

\section{Entropy convergence: effective measure complexity \label{sec:emc}}

Four magnitudes have been defined: $\zeta_\ell$, $h_\ell$, $\Delta H_\ell$ and $k_\ell$, which, in the limit of infinite strings, converge to the KS entropy, yet the way they converge is different.

Let $\eta_\ell$ be any of the mentioned magnitudes for finite $\ell$; then $\delta h^{(\eta)}_\ell=|\eta_\ell-h|$ is the deficit of the given magnitude with respect to its convergent value at time step $\ell$.  $\delta h^{(\eta)}_\ell$ is a magnitude of information, and its precise interpretation depends on the particular $\eta$. Regardless of its interpretation the sum of the $\delta h^{(\eta)}_\ell$ with $\ell$ can be taken as some sort of discrete integration \cite{crutchfieldRegularitiesUnseenRandomness2001}
\begin{equation}
E_{\eta}=\sum\limits_{\ell=1}^{\infty} \delta h^{(\eta)}_\ell, \label{eq:Eeta}
\end{equation}
and it's directly related to the convergence speed of the involved magnitude $\eta$. The larger the sum, the slower the convergence. It must be noticed that the series given by (\ref{eq:Eeta}) may not converge at all.

As is often the case, for an exponential decaying dependence of
\begin{equation}
\delta h^{(\eta)}_\ell=(\eta_0-h)\, 2^{-\nu \ell},
\end{equation}
$E_{\eta}$ will have the form
\begin{equation}
E_{\eta}=\frac{\eta_0-h}{2^{\nu}-1}.
\end{equation}
For a power law decay, $\delta h^{(\eta)}_\ell\propto \ell^{-\gamma}$, the sum (\ref{eq:Eeta}) converges only if $\gamma>1$, while for decaying rates between these two behaviours, of the form
\begin{equation}
\delta h^{(\eta)}_\ell\propto 2^{-\nu \ell^\beta},\;\;\;\; 0<\beta<1,
\end{equation}
it will converge provided that $\nu > 0$.

For all the $\zeta_\ell$, $h_\ell$, $\Delta H_\ell$ magnitudes, from the results in the previous sections, it follows that for a random sequence $E_\eta=0$. For a periodic sequence of minimal period $p\geq 2$, equations (\ref{eq:hlp}) and (\ref{eq:zetap}) give $\zeta_\ell=h_\ell=\log p/\ell$ for $\ell\geq p$ and $h=0$, so that, for $\eta_\ell=\zeta_\ell$ or $h_\ell$, $E_\eta$ contains the harmonic series $\log p\sum_{\ell\geq p}\ell^{-1}$ and does not converge, which is not a nice property to characterise a periodic sequence, as it does not distinguish among them. Yet, it is quite straightforward to see that in the case of $\Delta H_\ell$, for a periodic sequence of minimum period $p$,
\begin{equation}
E_{\Delta H}=\log p, \label{eq:ezeta}
\end{equation}
which makes it a much more interesting variable, and, in what follows, from all the $E_\eta$, we will be dealing only with $E_{\Delta H_\ell}$, the subscript will be dropped,
\begin{equation}
E=\sum\limits_{\ell=1}^{\infty} (\Delta H_\ell-h).\label{eq:E}
\end{equation}
$E$ was introduced by Grassberger \cite{grassbergerQuantitativeTheorySelfgenerated1986} under the name of effective measure complexity (EMC), but is also known as excess entropy \cite{crutchfieldRegularitiesUnseenRandomness2001}.

From the comment that was made after equation (\ref{eq:entgain1}), $E$ is the minimal amount of information that has to be stored in order to make optimal predictions under absolute efficiency \cite{grassbergerQuantitativeTheorySelfgenerated1986}. For sources where $E$ converges, the minimal memory required is finite, and such sources have been called finitary \cite{crutchfieldRegularitiesUnseenRandomness2001}.

Consider the bi-infinite string $u=u^-u^+$ partition into two halves $u^-$ and $u^+$, the first known as the past and the latter as the future. Take the $\ell$ last symbols from the past, $u^-_\ell$, and the first $\ell$ symbols from the future, $u^+_\ell$; the mutual information between the two finite strings is well defined
\begin{equation}
I[u^-_\ell, u^+_\ell]=H[u^+_\ell]-H[u^+_\ell|u^-_\ell].\label{eq:mutuali}
\end{equation}
The first term is equal to $H_\ell$, and the second term, using the chain rule, can be written as
\[
H[u^+_\ell|u^-_\ell]=\sum\limits_{i=0}^{\ell-1}H[u_i|u_{-\ell}u_{-\ell+1}\ldots u_{i-1}].
\]
Using both terms, the following expression follows,
\begin{equation}
I[u^-_\ell, u^+_\ell]=H_\ell-\sum\limits_{i=0}^{\ell-1}H[u_i|u_{-\ell}u_{-\ell+1}\ldots u_{i-1}]. \label{eq:mutualic}
\end{equation}
Consider the expression,
\[
\sum\limits_{i=1}^{\ell}[\Delta H_i-h]=H_\ell-\ell h
\]
In the limit $\ell\rightarrow \infty$, this last expression is equal to $E$,
but this is precisely the expression (\ref{eq:mutualic}) in the same limit; therefore, it has been proven that \cite{crutchfieldRegularitiesUnseenRandomness2001}
\begin{equation}
E=I[u^-, u^+].\label{eq:emutuali}
\end{equation}
The effective measure complexity is the mutual information between the two halves of the bi-infinite string $u$, further emphasising the interpretation given above. If convergence is uniform, then asymptotically \cite{crutchfieldRegularitiesUnseenRandomness2001}
\begin{equation}
H_\ell\sim E+\ell h,\label{eq:asymptH}
\end{equation}
for $\ell\gg1$.

As $E$ is the mutual information between the two halves of the bi-infinite string, it can be written
\begin{equation}
E=H(u^-)+H(u^+)-H(u).\label{eq:einfred}
\end{equation}
The effective measure complexity is then a measure of how much more information is needed to specify the sequence $u$ when it is split into two independent halves, $u^-$ and $u^+$, than to specify the single sequence $u$. This is consistent with the stated fact that, for a random sequence (and, indeed, for any i.i.d.\ sequence), $E=0$; any other (positive) value of $E$ signals correlation in the sequence's partition, and memory is then understood as a measure of such correlation.

Following from the above comment, there is something to be learned not only from the value of $E$, but also from the terms in the sum in equation (\ref{eq:E}). If for a finite $\ell_c$, $\Delta H_\ell$ converges to $h$, then correlations in the system can be found up to $\ell_c$. Even if that is not the case, the decreasing behaviour of the terms $\Delta H_\ell-h$ as $\ell$ increases is related to the decay of correlations in the system over time.

\section{Information distance\label{sec:dist}}

Algorithmic complexity provides an adequate framework for defining the informational distance between two strings \cite{bennet98,liSimilarityMetric2004}. What is sought is a measure of the minimal amount of information needed to compute one string from the other and vice versa. If this measure is to be considered a distance, apart from being non-negative, it must satisfy the identity axiom, that the distance from one string to itself of zero; the symmetry axiom, where the distance from string one to string two is equal to the distance from string two to string one; and the triangle inequality that states that the distance between two string is smaller or equal to the sum of the distance of both strings to a third one.

Let $w_1$ and $w_2$ be two strings. We define the information distance as the length of the shortest program that computes $w_1$ from $w_2$ as well as $w_2$ from $w_1$. It can be shown that if
\begin{equation}
K(w_1|w_2)>K(w_2|w_1), \label{eq:kk}
\end{equation}
Then the program that computes $w_1$ given $w_2$ can be made to contain the information needed to generate the program that computes $w_2$ given $w_1$ \cite{bennet98}. This last fact leads to defining the (absolute) informational distance as
\begin{equation}
d_\parallel(w_1,w_2)= \max\{K(w_1|w_2),K(w_2|w_1)\}. \label{eq:absd}
\end{equation}

The defined absolute distance is clearly non-negative and complies with the symmetric axiom, as well as the identity axiom up to a term $O(1/K(w_i))$, where $w_i$ is the string with the largest algorithmic complexity. In \cite{liSimilarityMetric2004}, it is shown that it also complies with the triangular inequality, also up to a term $O(1/K(w_i))$ where $w_i$ is now the string with the maximum Kolmogorov complexity of the three strings involved in the triangle inequality.

Zurek reasoned about the relation between algorithmic complexity and thermodynamic cost \cite{zurek89, zurek89a}: the minimum increase of thermodynamic entropy for a computation that turns string $w_1$ into $w_2$, without retaining memory of the input or of the performed steps, should be equal to $K(w_1|w_2)+O(1)$.  The demonstration is based on the Landauer principle that states that an erasure of a single bit implies a dissipation of energy of no less than $k_BT\ln 2$, where $k_BT$ is the Boltzmann product \cite{landauer61}. If $w_2$ contains all the information necessary to compute $w_1$, then the process is reversible, and there is no increase in thermodynamic entropy. On the contrary, if $w_1$ cannot be recovered from $w_2$, information has been lost, and thermodynamic entropy has increased. The absolute information distance can then be related to the maximum thermodynamic cost required to produce one string from the other, and vice versa.

The distance defined by (\ref{eq:absd}) encounters practical difficulties when comparing strings of different lengths. Two independent random strings of small length will have a smaller distance than two algorithmically related strings of much larger length, with just a small amount of dissimilar information. To avoid this dependence on the string length, the normalised informational distance has been defined as
\begin{equation}
d(w_1,w_2)= \frac{\max\{K(w_1|w_2),K(w_2|w_1)\}}{\max\{K(w_1),K(w_2)\}}. \label{eq:normd}
\end{equation}
A discussion can be made on whether the denominator in (\ref{eq:normd}) is the proper normalisation factor; the reader can follow Li and Vitanyi \cite{liSimilarityMetric2004} for a justification of the chosen factor.

From the definition of the conditional algorithmic complexity and the algorithmic mutual information, it follows
\begin{equation}
d(w_1,w_2)=1-\frac{K(w_1:w_2)}{K(w_2)}. \label{eq:infd}
\end{equation}
where  it has been assumed that $K(w_2)>K(w_1)$. From this last expression (\ref{eq:infd}), $1-d(w_1,w_2)$ is the amount of information (in bits) that is shared between both strings per bit of information of the string with the largest Kolmogorov complexity \cite{liSimilarityMetric2004}.

The information distance should not be confused with entropy-based divergences between probability distributions, such as the Kullback-Leibler divergence or the Jensen-Shannon divergence \cite{lin1991divergence}, whose square root is a metric \cite{endres2003new}. Applied to two sequences, such divergences compare their (empirical) word distributions, and therefore quantify whether both sequences share the same statistics, that is, whether they look like realisations of the same source. The information distance quantifies instead how much information one individual sequence carries about the other. Two independent realisations of the same source have a vanishing Jensen-Shannon divergence but an information distance close to one, while two strongly coupled sequences with different statistics can be far apart in the Jensen-Shannon sense and close in the sense of $d$. The information distance is thus the appropriate measure when coupling, rather than statistical similarity, is sought, as in the case studies of Section~\ref{sec:case}. In addition, the Kullback-Leibler divergence is neither symmetric nor satisfies the triangle inequality, and both divergences require estimating length-$\ell$ word distributions over $\alpha^\ell$ cells, with the undersampling problems discussed in Section~\ref{sec:flow} for the transfer entropy, whereas the estimate of $d$ introduced in Section~\ref{sec:est} requires no block length.

If all strings involved are of the same length, in the right member of $(\ref{eq:normd})$, the numerator and the denominator can be divided by the length and, in the limit of $\ell\rightarrow \infty$, the following expression follows
\begin{equation}
d(w_1,w_2)= \frac{\max\{h(w_1|w_2),h(w_2|w_1)\}}{\max\{h(w_1),h(w_2)\}}. \label{eq:hd}
\end{equation}
The practical importance of this last expression is that while the Kolmogorov complexity is uncomputable, the entropy density is not. To achieve this, we have had to consider strings of the same length and normalised distance. In spite of the uncomputable nature of the Kolmogorov complexity, it has been demonstrated before that the growth rate of algorithmic complexity can be estimated, notably by Zurek, who derived the Sackur-Tetrode equation for the entropy of an ideal gas from the Kolmogorov complexity rate \cite{zurek89a}.

Equation (\ref{eq:hd}) will be used for computing the information distance between two sequences in practical terms and is the mathematical basis for the normalised compression distance used by Li and Vitanyi \cite{liSimilarityMetric2004}.

Given a set of strings, the information distance between all pairs of strings allows us to build a distance matrix and, from there, an undirected weighted graph among the strings. Such a graph will be called an information graph. If a proper threshold is set, the distance matrix becomes a connectivity matrix, and the corresponding graph is unweighted and undirected.

\subsection{Flow of information\label{sec:flow}}

Considering the inequality (\ref{eq:kk}), the net flow of information between two strings can be defined by \cite{zurek89},
\begin{equation}
\Delta K=K(w_1|w_2)-K(w_2|w_1). \label{eq:flow}
\end{equation}
if $\Delta K >0$, then the information flows from $w_1$ to $w_2$, when $\Delta K<0$, the opposite is true. The net flow allows us to turn the information graph into a directed graph that shows the flow of information in a given system.

As only the sign of $\Delta K$ is of interest for the graph, the following variant can be used instead of equation (\ref{eq:flow}),
\begin{equation}
\Delta h=h(w_1|w_2)-h(w_2|w_1). \label{eq:flowh}
\end{equation}
When the sign is not enough and actual values are needed, the same argument used to derive (\ref{eq:hd}) can be used to justify (\ref{eq:flowh}), if both strings have the same length.

Flow of information as defined above is related but can not be identified with causality. Causality needs to account for the time in the flow of information between two strings. The concept of transfer entropy is then better suited for causality relations \cite{sch00}.

Transfer entropy was introduced by Schreiber \cite{sch00} to quantify directed, time-asymmetric dependence between two processes. For a target series $w_j$ and a source series $w_i$, sampled as $w_j(t)$ and $w_i(t)$, let $w_j^{(m)}(t)=\big(w_j(t),\dots,w_j(t{-}m{+}1)\big)$ denote the $m$-block past. The transfer entropy from $w_i$ to $w_j$ is
\begin{equation}
TE_{w_i\to w_j}=\sum P\big(w_j(t{+}1), w_j^{(m)}(t), w_i^{(m)}(t)\big)\,
\log\frac{P\big(w_j(t{+}1)\mid w_j^{(m)}(t),w_i^{(m)}(t)\big)}{P\big(w_j(t{+}1)\mid w_j^{(m)}(t)\big)},
\label{eq:te_schreiber}
\end{equation}

the sum over all realisations of its arguments. This is the conditional mutual information $I\big(w_j(t{+}1);w_i^{(m)}(t)\mid w_j^{(m)}(t)\big)=H\big(w_j(t{+}1)\mid w_j^{(m)}(t)\big)-H\big(w_j(t{+}1)\mid w_j^{(m)}(t),w_i^{(m)}(t)\big)$, the information the source past adds about the target's next symbol beyond the target's own past. It vanishes when that information is null and is in general asymmetric, $TE_{w_i\to w_j}\neq TE_{w_j\to w_i}$.

The direct estimation of (\ref{eq:te_schreiber}) is problematic; it requires the joint distribution over a space of dimension $2m+1$; for a symbolic source of alphabet size $\alpha$ this is a histogram of $\alpha^{2m+1}$ cells, so the sample size needed to populate it grows exponentially with the embedding order, and most cells remain empty or undersampled for realistic record lengths (for continuous data an additional partitioning or kernel step is required, introducing further free parameters). Plug-in entropy estimators are, moreover, systematically biased, and because the two terms of the difference live in spaces of different dimension their biases do not cancel, so that independent series can yield a spuriously positive value. The estimate is therefore sensitive to $m$, $\ell$, the embedding delay and the partition, and demands long stationary records \cite{res20}.

These difficulties motivate working with the per-symbol rate and estimating it without explicit density estimation. Taking the block pasts to the $m$-step past and normalising per symbol turns the block entropies of (\ref{eq:te_schreiber}) into entropy densities, and the transfer entropy rate from $w_i$ to $w_j$ becomes

\begin{equation}
TER_{w_i\to w_j}=h\big(w_j\mid w_j^{<m}\big)-h\big(w_j\mid w_j^{<m},w_i^{<m}\big),
\label{eq:te_rate}
\end{equation}
with $w^{<m}$ the $m$-step past. Equation (\ref{eq:te_rate}) is non-negative, vanishes when $w_i$'s past adds no information about $w_j$'s present, and is asymmetric under $i\leftrightarrow j$, thus encoding directionality \cite{sch00,res20}. It is the quantity estimated below with the Lempel-Ziv factorisation, where a same-dimension surrogate removes the residual bias that afflicts (\ref{eq:te_schreiber}).

\paragraph*{Transfer entropy rate criterion.} At finite $m$ and finite sequence length, the two densities in (\ref{eq:te_rate}) carry an embedding-dimension bias that does not cancel exactly. We remove it with a surrogate of the same embedding dimension: let $_\mathrm{S}w_i$ be $w_i$ with its past block time-permuted, which destroys its temporal relation to $w_j$ while preserving its marginal statistics and the embedding size. The surrogate-corrected transfer entropy rate is
\begin{equation}
\widehat{\mathrm{TER}}_{i\to j}=\big\langle \widehat h\!\left(w_j\mid w_j^{<m}, {_\mathrm{S}w_i^{<m}}\right)\big\rangle_{\Gamma}-\widehat h\!\left(w_j\mid w_j^{<m},w_i^{<m}\right),
\label{eq:ter}
\end{equation}
where $\langle\cdot\rangle_{\Gamma}$ averages over $\Gamma$ surrogates. The surrogate reproduces the bias of the real estimate, so $\widehat{\mathrm{TER}}_{i\to j}\ge 0$ measures the genuine reduction in the entropy density of $w_j$ due to $w_i$'s real, as opposed to scrambled, past.

\section{Lempel-Ziv factorisation\label{sec:lz}}

In this section, we follow \cite{lempelComplexityFiniteSequences1976} and \cite{ziv78}.

Given a string $w$, the set of all distinct words of $w$ is called the vocabulary of $w$ and is denoted by $\mathcal{V}(w)$. For example, $w=10011$ has vocabulary $\mathcal{V}(w) = \{ \Lambda, 0, 1, 10, 00, 01, 11, 100, 001, 011, 1001, 0011, 10011 \}$.

A sequence $w$ can be partitioned into factors, that is, the concatenation of non-overlapping words,
\[
w=F_1F_2\ldots F_{n_F},
\]
such that  the word $F_i(w)=w_{a_i}^{a_{i+1}-1}$, where $a_i$ is the position in $w$ at which the factor $F_i$ begins.

A history of the finite string $w$ is a factorisation such that
\begin{equation}
F_i\pi\in \mathcal{V}(w_{0}^{a_{i+1}-1}\pi^2)
\end{equation}
except, perhaps, for the first factor $F_1$. $\pi$ is the drop operator defined earlier. For example, for $w=1\cdot0\cdot01\cdot11\cdot010\cdot110\cdot00\cdot$, a history is described by the factors delimited by the dots.

Lempel-Ziv's approach to sequence complexity derives from considering the unique history of $w$, which is the result of a recursive factor concatenation process starting from a prefix factor $ F_1 = w_0 \in \mathcal{A}$ that allows for innovation. Innovation, in this context, refers to the fact that each factor $F_{i+1}$ adds a word to the vocabulary of the growing sequence $\mathcal{V}(w_{0}^{a_{i+1}-1})$. Such a process is called an exhaustive production process and complies with,
\begin{equation}
F_i\notin \mathcal{V}(w_{0}^{a_{i+1}-1}\pi), \label{eq:ehistory}
\end{equation}
for all factors in the history of $w$, except, perhaps, the last factor of the factorisation.

For example, the exhaustive factorisation of the above example $w=10011101011000$, will be $1\cdot0\cdot01\cdot110\cdot101\cdot1000$. An exhaustive factorisation is also called a Lempel-Ziv factorisation, $LZ(w)$. The Lempel-Ziv factorisation of a string is unique.

The Lempel-Ziv complexity of a sequence (from now on LZ-complexity) is defined as the number of factors in the exhaustive factorisation of the sequence,
\begin{equation}
C_{LZ}(w)=|LZ(w)|.\label{eq: CLZ}
\end{equation}

Lempel-Ziv complexity is a measure of a sequence's innovativeness. Each factor in the Lempel-Ziv history corresponds to a single innovation (at the ending symbol of the factor). The larger the number of factors in $LZ(w)$, therefore, the larger the innovation process that produces the string. From this interpretation, it is clear that an appropriate normalisation of $C_{LZ}$ can be taken, related to the innovation per symbol of the sequence.

\subsection{Properties of the LZ-factorisation}

\begin{enumerate}
\item The exhaustive history of a sequence is the smallest possible history factorisation.

\textbf{Demonstration:} See appendix \ref{app:ex}.

\item For every string $w$ with $N=|w|$,
\begin{equation}
C_{LZ}(w)<\frac{N}{(1-\epsilon_N)\log N}, \label{eq:clzineq}
\end{equation}
where
\begin{equation}
\epsilon_N=2\frac{1+\log\log (\alpha N)}{\log N}.
\end{equation}

\textbf{Demonstration:} See appendix \ref{app:bound}.

\item Let $w$ be a string of length $N\gg 1$ emitted by an ergodic source of entropy density $h$. Then,
\begin{equation}
C_{LZ}(w)\leq N h/\log N.\label{eq:clzhN}
\end{equation}

\textbf{Demonstration:} See appendix \ref{app:length}.

\item Given the string formed by concatenating the sequences $w_1$ and $w_2$, $C_{LZ}(w_1w_2)\leq C_{LZ}(w_1)+C_{LZ}(w_2)$.

\textbf{Demonstration \cite{lempelComplexityFiniteSequences1976}:} Consider a history of $w_1w_2$, such that $F(w_1w_2)=LZ(w_1)\cdot LZ(w_2)$, that is the concatenation of the LZ-factorisation of $w_1$ and $w_2$. From the first property $C_{LZ}(w_1w_2)\leq|F(w_1w_2)|=C_{LZ}(w_1)+C_{LZ}(w_2)$ which proves the property.
\end{enumerate}

\subsection{Shannon entropy and LZ-factorisation}

Define the normalized Lempel-Ziv complexity as \cite{lempelComplexityFiniteSequences1976}
\begin{equation}
c_{LZ}(u)=\limsup\limits_{N\rightarrow \infty} \frac{C_{LZ}(u)}{N/\log N}. \label{eq:clzn}
\end{equation}
From property 3, follows
\begin{equation}
c_{LZ}\leq h.\label{eq:ineqclzh}
\end{equation}
On the other hand, in \cite{ziv78} it is shown that $C_{LZ}$ is the basis for a lossless finite-state encoder, so according to the converse theorem of information theory \cite{coverElementsInformationTheory2006}, it follows that
\begin{equation}
\liminf\limits_{N\rightarrow\infty} c_{LZ}\geq h. \label{eq:ineqclzh1}
\end{equation}
Combining inequalities (\ref{eq:ineqclzh}) and (\ref{eq:ineqclzh1}), the following theorem \cite{ziv78} follows (theorem 7 of Ziv):
\begin{equation}
P\left[|c_{LZ}(u)-h|>\epsilon\right]=0, \label{eq:ziv}
\end{equation}
where $\epsilon>0$ is an arbitrary constant.

\section{Estimation of entropic magnitudes using Lempel-Ziv factorisation\label{sec:est}}

\subsection{Entropy density}

Theorem $7$ of Ziv (\ref{eq:ziv}) is the basis for using the Lempel-Ziv factorisation as an estimate of the entropy rate. The normalized Lempel-Ziv complexity has to be computed for a necessarily finite-length sequence, and therefore some error is unavoidable in the estimate. The error is related to $\epsilon_N$, but, as shown in \cite{lesne09}, Lempel-Ziv estimation of $h$ is quite efficient at approximating the density rate even when relatively short sequences are used. This may come as a surprise, taking into account that $\epsilon_N$ is a very slowly decaying function achieving a value of around $0.4$ for a binary alphabet with strings as long as $1$ million symbols, and only falls to $0.3$ for a string length of $10^{9}$.

Consider $u$ an infinite string that starts at $i=0$. If $u_0^N$ is a prefix of length $N$ of $u$, then, by construction, $C_{LZ}(u_0^N)$ is a non-decreasing function of $N$. For very small sequences ($N=0,1,2,3\ldots$), one should expect, in general, that $C_{LZ}(u_0^N)<N/\log N$ does not need to be true. As $N$ increases, the function $N/\log N$ keeps growing while redundant subsequences start to appear in the string and the $C_{LZ}(u_0^N)$ starts to grow more slowly with intervals of plateau. The more redundant the string in terms of patterns, the slower the growth rate. It should be expected that the fastest-growing behaviour for $C_{LZ}(u_0^N)$ should be achieved for a random string.

\begin{figure}[!ht]
\centering
\includegraphics*[scale=0.7]{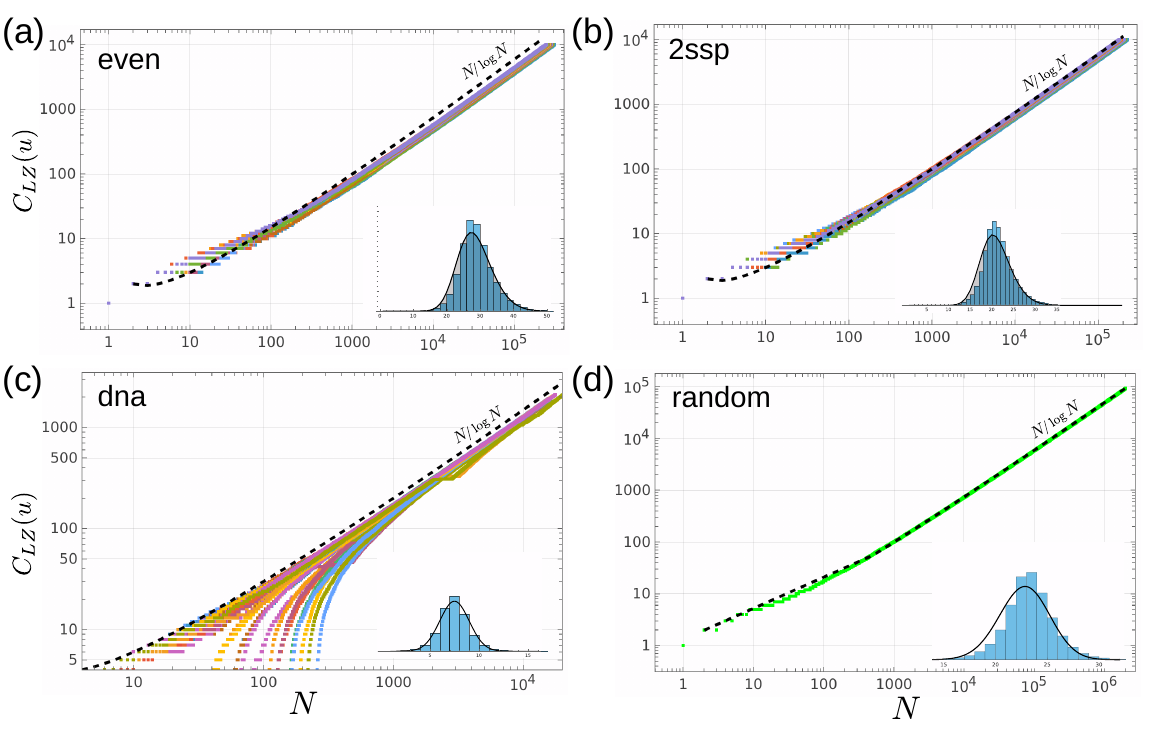}
\caption{\textbf{Lempel-Ziv complexity and the asymptotic behaviour.} The four plots show the Lempel-Ziv complexity as a function of the string length $N$, compared with the same curve for the asymptotic behaviour $N/\log N$ (Dashed black line). (a): even and (b): 2ssp correspond to finite-state automata defined in Appendix \ref{app:fsa}. The first is the even process, which is controlled by a transition probability. Each colour corresponds to a different value of the transition probability. The second one (2ssp) is for the two-state symmetric process, which is also governed by a transition probability. (c): dna pertains to an experimental example, the DNA sequences of different mammals were used to factorise and estimate the complexity value. Finally, (d): random, stands for a random string. In all cases, the LZ-complexity is upper-bounded asymptotically by the asymptotic function as expected. What is interesting is that for large string size, the behaviour with the string length settles for an ansatz as described by equation (\ref{eq:clza}). For short string lengths, all examples except the random case deviate from the asymptotic curve, which can be significant, as shown in the DNA data. The inset in each plot is the histogram, and the closer Normal curve of the distribution of factor lengths in the Lempel-Ziv factorisation.
}\label{fig:fact}
\end{figure}

This is exemplified in Figure \ref{fig:fact} where the $C_{LZ}(u_0^N)$ for four different symbol sources are shown as a function of the string length $N$. Two sources are unifilar finite-state automata, commonly discussed when analysing entropic measures \cite{crutchfieldRegularitiesUnseenRandomness2001} (see Appendix \ref{app:fsa}). A third source is an experimental DNA sequence for mammals, and the fourth source is a random string. For the finite-state sources, the described behaviour can be readily observed, where for small sequence length $C_{LZ}(u_0^N)$ values are above the asymptotic $N/\log N$ function, but as $N$ increases, it falls below. In the DNA case, a number of sequences start with a growth rate for $C_{LZ}(u_0^N)$ well below the asymptotic curve and the pace of growth increases with $N$. Finally, for the random string, the $C_{LZ}(u_0^N)$ almost coincides with $N/\log N$ for all the $N$ range. One interesting feature of all four examples is that for large $N$ values, $C_{LZ}(u_0^N)$ settles into a law
\begin{equation}
C_{LZ}(u_0^N)=h \frac{N}{\log N}-\kappa,\label{eq:clza}
\end{equation}
with $\kappa$ a constant. From this last equation
\begin{equation}
c_{LZ}(N)=h-\frac{\kappa}{N/\log N}.
\end{equation}
The estimation of the entropy density value from the $c_{LZ}(N)$ of a finite sequence is underestimated by an amount $\kappa\log N/N$. One could be tempted to estimate $h$ from the fit of $C_{LZ}(u_0^N)$ vs $N$ for large $N$ values, but one must be warned that the slow convergence of $c_{LZ}(N)$ implies that the fit will be highly sensitive to the smallest value $N_0=N$ chosen to start the fit.

Amig\'o and Kennel have obtained an expression for the variance in the estimation of $h$ by the normalized Lempel-Ziv complexity \cite{amigoVarianceEstimatorsLempelZiv2006}. Let $\lambda_k$ be the length of the $k$ factor in the Lempel-Ziv factorisation of a string. It follows
\begin{equation}
N=\sum\limits_{k=1}^{C_{LZ}}\lambda_k, \label{eq:amigo1}
\end{equation}
where $N$ is the length of the string.

A normal distribution $\mathcal{N}(\mu,\sigma^2)$ of the length of the factors, with mean $\mu$ and variance $\sigma^2$, is assumed. If $n_F$ factors are considered to be drawn from a normal distribution, then the length $N$ will be a random variable with expected value $n_F\mu$ and variance $n_F\sigma^2$. $N$ can then be written as
\begin{equation}
N=n_F\mu+(n_F^{1/2}\sigma)Z, \label{eq:amigo2}
\end{equation}
where $Z$ is a random variable drawn from a normal distribution $\mathcal{N}(0,1)$. Let us reverse the point of view: fix $N$ and consider $n_F$ a random variable. In such a case, the estimated entropy density is
\begin{equation}
\widehat{h}=n_F \log N/N\label{eq:amigo3}
\end{equation}
and its standard deviation
\begin{equation}
\sigma_{\widehat{h}}=\sigma_{n_F} \log N/N.\label{eq:amigo4}
\end{equation}
From equation (\ref{eq:amigo2}),
\begin{equation}
Z=\frac{N-n_F\mu}{n_F^{1/2}\sigma}.\label{eq:amigo5}
\end{equation}
Propagating errors in this last equation
\begin{equation}
\sigma_Z=\left(\frac{\mathrm{d}Z}{\mathrm{d}n_F}\right)\sigma_{n_F},
\end{equation}
but $\sigma_Z=1$ by construction, so, using equation (\ref{eq:amigo5})
\begin{equation}
\sigma_{n_F}=\left|\left( \frac{\mathrm{d}Z}{\mathrm{d}n_F}\right)\right|^{-1}=\frac{2n_F^{3/2}\sigma}{n_F\mu+N}.\label{eq:amigo6}
\end{equation}
The best estimate for $n_F$ is $C_{LZ}$, taking $\mu=N/C_{LZ}$, the following expression follows
\begin{equation}
\sigma_{\widehat{h}}=\sigma\frac{c_{LZ}(N)^{3/2}}{\sqrt{N\log N}},\label{eq:amigo7}
\end{equation}
where equation (\ref{eq:amigo3}) and (\ref{eq:amigo4}) were used. The reader must recall that $\sigma^2$ is the variance in the length of factors that can be obtained from the factorisation. Expression (\ref{eq:amigo7}) shows that the larger $c_{LZ}(N)$, the larger the error in its estimation.

Following the same procedure, we include the variance for a Poisson distribution in the factor length, resulting in
\begin{equation}
\sigma_{\widehat{h}}=\frac{c_{LZ}(N)}{N}.\label{eq:amigo8}
\end{equation}

The histogram over the factor length for each of the described processes can be well approximated by a Normal distribution (see insets in Figure \ref{fig:fact}), making the use of equation (\ref{eq:amigo7}) as an estimate of the error in $\widehat{h}$ valid.

\begin{figure}[!t]
\centering
\includegraphics*[scale=0.6]{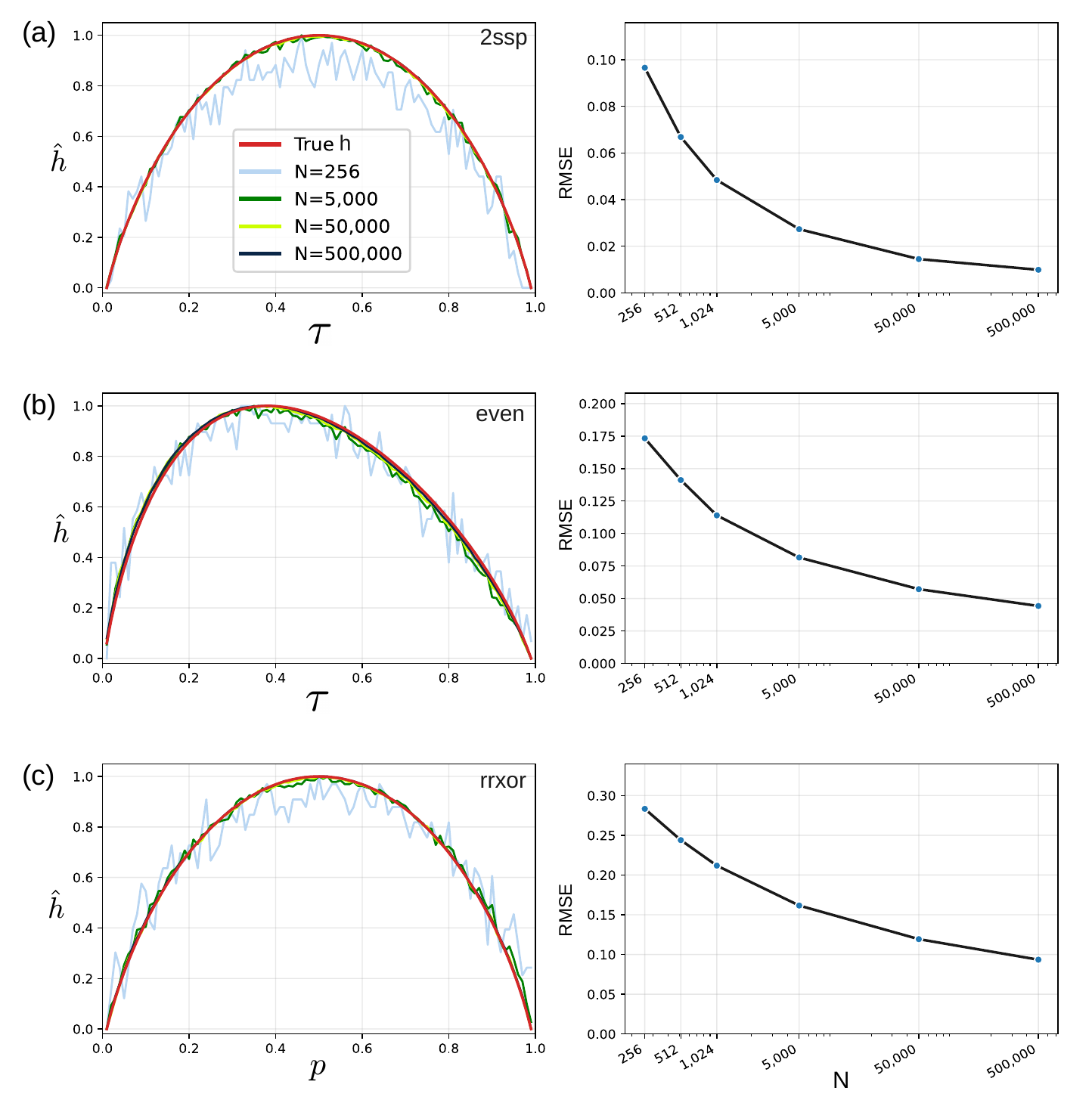}
\caption{\textbf{Convergence of the Lempel-Ziv entropy-density estimate with sequence length.} Each row corresponds to one Hidden Markov Model process: (a) the two-state symmetric process (2SSP, top), (b) the even process (middle), and (c) RRXOR (bottom), swept across their full control-parameter range. \emph{Left column:} the Lempel-Ziv estimate $\widehat h$ at four sequence lengths ($N=256$; $5{,}000$; $50{,}000$; $500{,}000$) overlaid on the true entropy density (red), each curve independently normalised between 0 and 1, $(h-\min)/(\max-\min)$, to compare shape convergence independently of each process's own entropy range. \emph{Right column:} root-mean-square error of the raw, un-normalised $\widehat h$ against the true entropy density, computed over the same parameter sweep, as a function of sequence length (log scale). In both columns, entropy density shows no qualitative convergence failure for any of the three sources, only a shrinking quantitative bias.}\label{fig:hFSA}
\end{figure}

The left column of Figure \ref{fig:hFSA} are the plot of the estimated entropy density $\widehat{h}$ from the Lempel-Ziv factorisation for different sequence lengths in the already presented finite-state automata corresponding to the symmetric 2-state process (a), the even process (b) and the random-random-xor process (c) for all values of their control parameters. First, the estimate follows the general trend even for sequence lengths as short as $256$ symbols, and it's almost indistinguishable from the exact entropy density $h$ for sequence lengths above $10^3$ symbols. For the two-state symmetric process, the estimate is even better as it behaves almost identically to $h$ for sequences of $256$ symbols and above.

\subsection{Effective measure complexity}

Estimating the effective measure complexity is non-trivial. As a measure
of correlation at all length scales, or the memory of the system, its
precise evaluation implies, in the general case, sums over infinite terms
\cite{grassbergerQuantitativeTheorySelfgenerated1986} and convergence is
usually slower than estimating entropy density. This makes any estimation
more cumbersome than estimating the entropy density, both analytically
and numerically, and less reliable for smaller sequence length.

One of the approaches followed, avoiding direct data estimation, is to
try to build the minimal computational machine able to statistically
reproduce the data. This approach was pioneered by Grassberger
\cite{grassbergerQuantitativeTheorySelfgenerated1986} and further
developed by Crutchfield and Feldman \cite{crutchfieldRegularitiesUnseenRandomness2001}.
In practice, such an approach has not gone beyond finite-state
approximations of the data, and even then, the reconstruction procedure
has been applied only to simple sources.

Section~\ref{sec:emc} singled out the entropy gain $\Delta H_\ell$,
Eqs.~\eqref{eq:entgain}-\eqref{eq:entgain1}, as the preferred building
block for the effective measure complexity: unlike $h_\ell$, $\zeta_\ell$ or
$k_\ell$, it converges to $h$ monotonically and, by Appendix~\ref{app:convergence},
at least as fast as $h_\ell$ does, $h_\ell(u)\geq \Delta H_\ell(u) \geq h$.

The corresponding estimator is
\begin{equation}
\Delta \widehat H_\ell \;\equiv\; \widehat H_\ell - \widehat H_{\ell-1},
\label{eq:DeltaHlhatCS}
\end{equation}
and, inspired by Eq.~\eqref{eq:E}, the effective measure complexity is
still estimated as
\begin{equation}
\widehat E(n) \;=\; \sum_{\ell=1}^{n}\left(\Delta \widehat H_\ell - \widehat h\right)
\;=\; \widehat H_n - n\,\widehat h,
\label{eq:EhatCS}
\end{equation}
with $\widehat h$ the entropy density of the original sequence
estimated exactly as before, from the Lempel-Ziv factorisation. Only $\Delta \widehat H_\ell$'s have to be estimated, which involves the estimate of the
block entropy $H_\ell(u)$ -- the Shannon block entropy of length-$\ell$. 

Melchert and Hartmann's approach \cite{melchertAnalysisPhaseTransition2015} obtains this indirectly: partition $u$ into non-overlapping blocks of length $\ell$, shuffle the blocks into a surrogate sequence $u^{RS(\ell)}$ (destroying correlations
above scale $\ell$ while preserving the length-$\ell$ block distribution
exactly), and read $H_\ell(u)$ off $u^{RS(\ell)}$'s own Lempel-Ziv complexity
via Ziv's theorem.

This is a consistent estimator, but it inherits a slow-convergence property of the Lempel-Ziv complexity itself: once the number of distinct length-$\ell$ blocks $\#_\ell$ actually observed becomes comparable to the number of blocks $m_\ell$ available to estimate their distribution from, the normalised Lempel-Ziv complexity underestimates entropy density by an amount that grows, not shrinks, with $\ell$. Under the non-overlapping partition above, $m_\ell = N/\ell$ falls linearly as $\ell$ grows while $\#_\ell$ can grow as fast as $\min(m_\ell,\alpha^{\ell})$ so this regime is reached quickly: a controlled test with $N=5\times10^5$ i.i.d.\ draws over $\alpha=28{,}256$ equally likely categories -- comparable to the $\#_\ell$ vs.\ $m_\ell$ balance reached at moderate $\ell$ in the surrogate construction above -- recovers only $55$-$80\%$ of the true entropy via Lempel-Ziv, depending on the exact draw.

To avoid the convergence problem, the surrogate-and-Lempel-Ziv step is replaced with a direct, coverage-adjusted estimate of $H_\ell(u)$ from the block frequency histogram of $u$ itself, following Chao and Shen \cite{chaoNonparametricEstimationShannon2003}. Details are given in the Appendix \ref{app:block}.

\subsubsection{Choice of $n$}
\label{sec:choice-of-n}

It is left to define a convenient limit $n$ for the sum in
Eq.~\eqref{eq:EhatCS}. For a sequence of length $N$, the maximum $n$ that
can be examined with reliable statistical significance is inherently
limited. A common estimate derives from assuming the sequence contains
roughly $N/n$ independent $n$-words, yielding the condition $N \geq
n\,\alpha^n$. This approximation is only accurate for independently and
identically distributed sequences, where any dependence among $n$-words
arises solely from their overlapping nature -- the same caveat that
applies, one level down, to the sliding-window histogram of
Eq.~\eqref{eq:HlhatCS} itself: consecutive windows share $\ell-1$ symbols
and are therefore not independent draws, even though there are $N-\ell+1$
of them. Sliding rather than partitioning changes how much of the
sequence's information content is \emph{sampled} at each $\ell$, easing the
severely-undersampled regime identified above; it does not change how
much information content the sequence \emph{has} at that $\ell$, so it does
not relax the bound below.

Lesne et al.\ \cite{lesne09} considered more realistic scenarios in which
intrinsic correlations within the sequence also influence the $n$-words
and must be accounted for. By interpreting the entropy rate $h$ as the
average amount of new independent information contributed per observed
symbol, the nominal sequence length can be replaced with an effective
length $N_{\mathrm{eff}} \approx N h$. This adjustment
incorporates the effect of internal correlations. Consequently, the
constraint on $n$ becomes more restrictive,
\begin{equation}
n\,\alpha^n \;\leq\; N h.
\label{eq:nbound}
\end{equation}

\begin{figure}[!t]
\centering
\includegraphics*[scale=0.7]{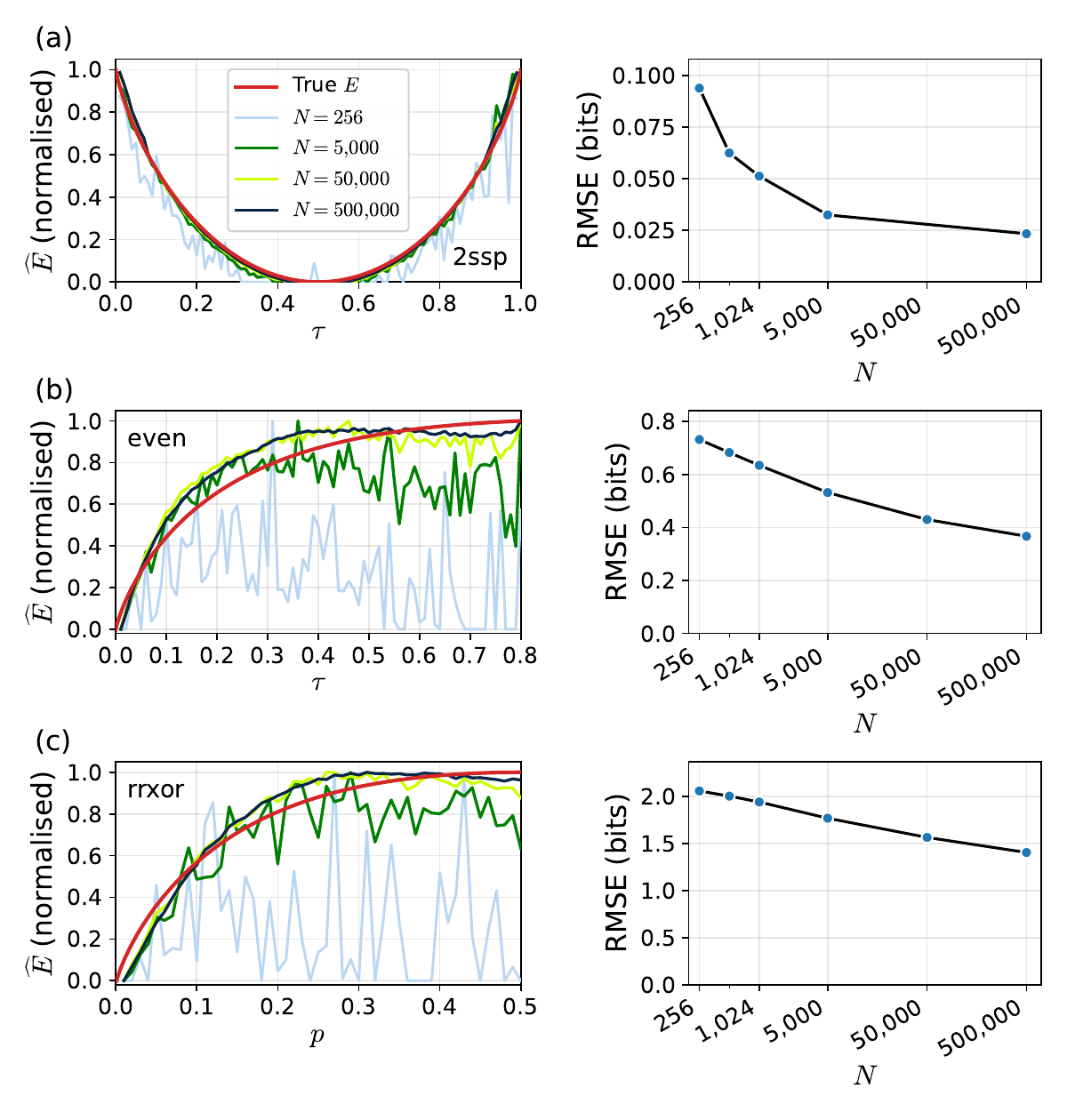}
\caption{\textbf{Estimation of the effective measure complexity.} Convergence of the effective measure complexity estimate across three Hidden Markov Model processes. Each row corresponds to one source: (a) the two-state symmetric process (2SSP, top), (b) the even process (middle), and (c) RRXOR (bottom). The control parameter $\tau$ is the transition probability of the two-state symmetric and even processes, and $p$ is the probability of a $1$ in the two free symbols of each RRXOR block (Appendix~\ref{app:fsa}, Fig.~\ref{fig:fsa}). Left column: the estimate $\widehat E$ at four sequence lengths ($N=256$; $5{,}000$; $50{,}000$; $500{,}000$) overlaid on the true effective measure complexity (red), each curve independently min-max normalized, $(E-\min)/(\max-\min)$, to compare shape convergence on a common scale. Right column: root-mean-square error of the raw, un-normalized $\widehat E$ against the true curve, over the same restricted range, as a function of sequence length. 2SSP converges cleanly to a low error floor by $N=50{,}000$; the even process and RRXOR both improve steadily but substantially more slowly, and neither has reached a comparable floor even at $N=500{,}000$, reflecting the longer block-size resolution their correlation structure demands relative to what $\widehat H_\ell$ can reliably support at this sample size.
}\label{fig:hEevensymm}
\end{figure}

\begin{figure}[!t]
\centering
\includegraphics*[scale=0.7]{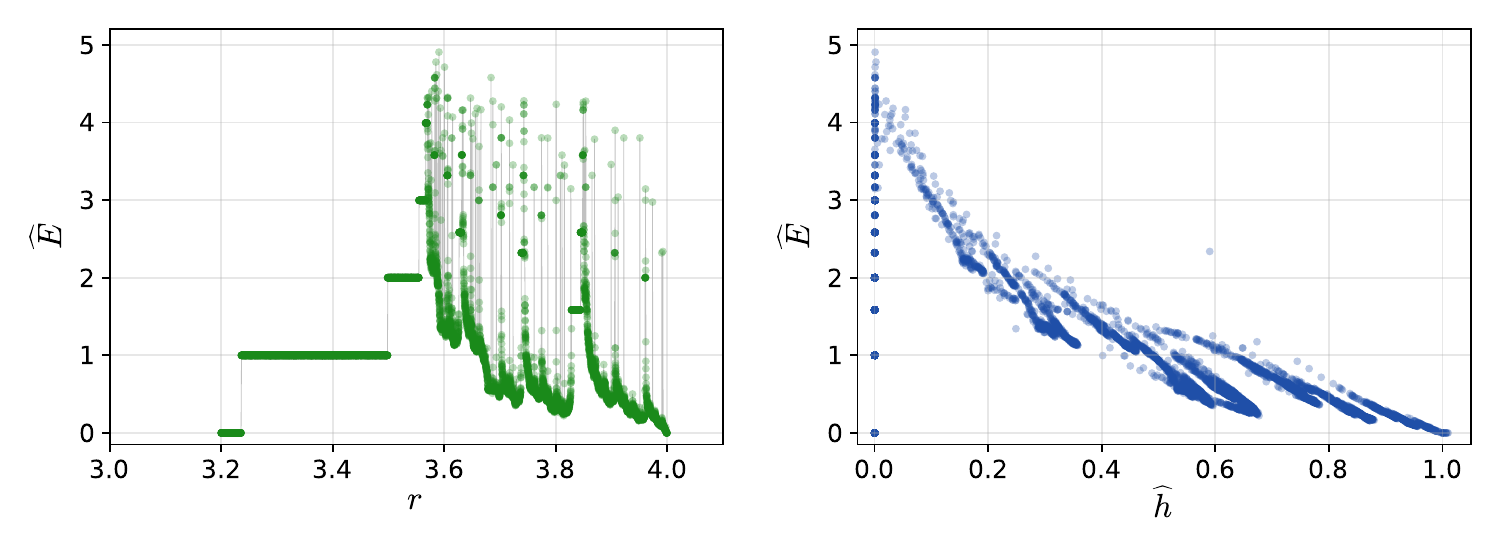}
\caption{\textbf{Estimation of the effective measure complexity in the logistic map.} The effective measure complexity estimated for the logistic map. (left) $\widehat{E}$ as a function of the control parameter $r$; (right) The complexity-entropy map. Both plots can be compared to figures 1 and 2 of \cite{feldman08}. 
}\label{fig:logistic}
\end{figure}

Claiming that the estimated $\widehat E$ equals the effective measure
complexity would be too strong, but its behaviour resembles that of $E$, as shown in Fig. \ref{fig:hEevensymm}. In Figure \ref{fig:logistic} the estimated effective measure complexity for the logistic map is shown (left), together with the complexity entropy map (right), which the reader can compare to the corresponding plots of Fig. 1 and 2, in \cite{feldman08}. The estimated $\widehat E$ captures properties of the sequence similar to those of the effective measure
complexity, with the correspondence expected to be closest for sources
whose correlation structure decays uniformly across their parameter
space, and to require correspondingly larger $n$ --- and hence longer
sequences, per Eq.~(\ref{eq:nbound}) --- for sources whose
synchronisation time varies strongly with the source's own control
parameters. 

\subsection{Information distance}

Following expression (\ref{eq:hd}), the conditional entropy can be written as $h(w_1|w_2)=h(w_2w_1)-h(w_2)$ which leads to the expression
\begin{equation}
\widehat{d}(w_1,w_2)= \frac{\max\{\widehat{h}(w_2w_1)-\widehat{h}(w_2),\widehat{h}(w_1w_2)-\widehat{h}(w_1)\}}{\max\{\widehat{h}(w_1),\widehat{h}(w_2)\}},\label{eq:dclz}
\end{equation}
for estimating the informational distance. $w_2w_1$ is taken as the ordered concatenation of sequences $w_2$ and $w_1$. The entropy densities $\widehat{h}$  involved in equation (\ref{eq:dclz}) can be estimated by the Lempel-Ziv factorisation as explained above. It must be noted that strictly speaking, $K(w_1|w_2)\approx K(w_2w_1)-K(w_2)$ as there is a logarithmic correction which becomes negligible for long enough sequences, and therefore the distance computed using (\ref{eq:dclz}) must not be for too short strings. However, if all strings studied are of the same length, then, without regard to the sequence's size, the distance can be used as a relative measure.

Following the same idea but starting with equation (\ref{eq:flowh}), the information flow can be estimated as
\begin{equation}
\Delta \widehat{h}=\widehat{h}(w_2w_1)-\widehat{h}(w_1w_2)+\widehat{h}(w_1)-\widehat{h}(w_2).\label{eq:flowhclz}
\end{equation}
$\Delta \widehat{h}>0$ means that information flows from $w_1$ to $w_2$, negative value will signal the opposite flow.

It must be clear what the distance expressed by equation (\ref{eq:dclz}) measures when $\widehat{h}$ is estimated by the Lempel-Ziv complexity, as compared to the more general definition using Kolmogorov complexity given by equation (\ref{eq:normd}). The informational distance, as defined by Kolmogorov complexity, pertains to how hard it is, algorithmically, to go from one sequence to another; ‘‘hard’’ here is measured by the length of the algorithm. In this regard, the informational distance is not a Hamming type of distance, it does not measure pairwise, how different are two strings and actually, a string of symbols and its negation will have (near to) zero informational distance, while the normalized Hamming distance will be $1$, as can be seen in the following example, where the string $w$ is the negation of $\tilde{w}$ and the Hamming field $\mathcal{H}$ is composed of only one’s:
\[
\begin{array}{l}
w=10110010011101010111100000\\
\tilde{w}=01001101100010101000011111\\
\mathcal{H}=11111111111111111111111111.
\end{array}
\]
However, a small algorithm of the type shown in Algorithm  \ref{alg:neg} will do the job, yielding a near-zero information distance.
\begin{algorithm}[H]
\begin{algorithmic}[1]
\For{$i$ from $1$ to $N$}
\State $\tilde{w}_i = w_i \otimes 1$
\EndFor
\State Print $\tilde{w}_i$
\end{algorithmic}
\caption{Computing the negated string $\tilde{w}$ from string $w$. The XOR operator is represented by $\otimes$}
\label{alg:neg}
\end{algorithm}

The same follows for a string of symbols and its reverse, where pairwise distance can be near one, and the Kolmogorov complexity would be, in all cases, (near) zero.

\begin{figure}[!ht]
\centering
\includegraphics*[scale=0.5]{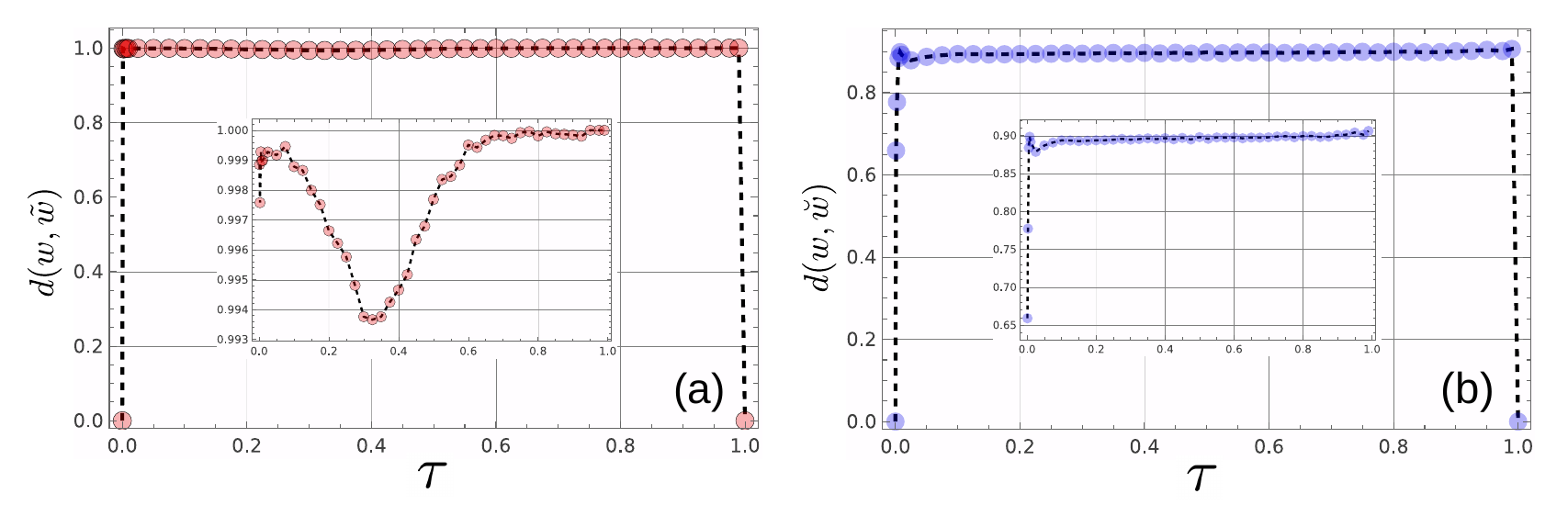}
\caption{\textbf{Estimation of information distance by Lempel-Ziv complexity}. Both plots show the information distance between the output string of the even process and a surrogate string for the whole range of transition probabilities. (a) The information distance between the string $w$ and its negated string $\tilde{w}$($=w\otimes 1$, where $\otimes$ is the XOR logical operator). (b) The information distance between the string $w$ and the reversed string $\breve{w}$. The inset shows a detailed scale of the same plot. In both cases, the distance is nearly $1$ for all transition probabilities, except for $\tau=0,1$.
}\label{fig:evend}
\end{figure}

When changing to a distance based on the definition (\ref{eq:dclz}), where $\widehat{h}$ is estimated by the Lempel-Ziv complexity, it must be understood that Lempel-Ziv factorisation is a pattern-driven procedure; it grows as new patterns emerge in the sequence. In this sense, it is also not a pairwise distance, but it is not all-encompassing as the Kolmogorov distance. The Lempel-Ziv-based distance can also be near 1 for two strings, whereas the Kolmogorov-based distance renders them close.

Figure \ref{fig:evend} shows, on the left, the plot of $\widehat{d}$ using Lempel-Ziv of the sequence of symbols produced by the even process for different values of the transition probability $\tau$ and the negated string, while on the right, a similar plot with the distance between the process string and its reverse. Except for the values at $\tau=0,1$, $\widehat{d}$ is near one in all other cases. This implies that when using Lempel-Ziv to compute the distance between two sequences, care must be taken to include, when necessary, simple operations that make sense for the process being analysed but are not captured by the Lempel-Ziv factorisation.

Let us clarify this last statement. We are not interested in abstract mathematical sources but rather in physical processes that can be considered sources of symbols (e.g., DNA, RNA, or the discretisation of a time signal). The physical laws governing the process will ultimately define a grammar for the source and the operations that can be regarded as within the ‘‘natural order’’ of the analysed system. It is within that set of rules that possible operations must be sought that can lead to a sequence transformation that could be missed by the Lempel-Ziv factorisation. Fortunately, this set of rules is almost always simple operations, and there is no need to seek involved ‘‘non-natural’’ transformations of the symbol sequence to compare with.

In most cases, a simple procedure as follows will suffice for calculating the distance between two strings $w_1$ and $w_2$:
\begin{enumerate}
\item Compute the distance $\widehat{d}$ between $w_1$ and $w_2$.
\item Compute the distance $\widehat{d}$ between $w_1$ and the reverse of $w_2$.
\item Compute the distance $\widehat{d}$ between $w_1$ and some well defined negation of $w_2$.
\item Take the distance as the smallest one among the three calculated distances.
\end{enumerate}

\subsection{Transfer entropy}

The conditional entropy densities in (\ref{eq:ter}) are estimated with the Lempel-Ziv factorisation over an extended alphabet, with the target and conditioner roles assigned as required by each term. Let $s_i(t)$ be the mean-binarised sequence of $w_i$ ($\alpha=2$). For a target series with present $s_i(t)$ and a conditioner $s_j$, form at each $t=m,\dots,T-1$ the state vector
\begin{equation}
v_t=\big(\underbrace{s_j(t{-}m),\dots,s_j(t{-}1)}_{\text{conditioner past}},\;\underbrace{s_i(t{-}m),\dots,s_i(t{-}1)}_{\text{target past}},\;\underbrace{s_i(t)}_{\text{target present}}\big)\in\{0,\dots,\alpha{-}1\}^{M},\quad M=2m+1,
\label{eq:statevec}
\end{equation}
and it collapses to a single symbol over an alphabet of size $\tilde\alpha=\alpha^{M}$ by the encoding
\begin{equation}
z_t=\sum_{r=0}^{M-1}\alpha^{r}\,v_t[r]\in\{0,\dots,\tilde\alpha-1\}.
\label{eq:encode}
\end{equation}
Applying the entropy-density estimate of Eq.~(\ref{eq:amigo3}) to the resulting sequence $z_m^{T-1}$ of length $N=T-m$, with the logarithm taken in base $\tilde\alpha$, gives
\begin{equation}
\widehat h(w_i\mid w_j)=\frac{C_{LZ}\!\left(z_m^{T-1}\right)\log_{\tilde\alpha} N}{N}.
\label{eq:embed_h}
\end{equation}
The marginal density $\widehat h(w_i)$ follows by dropping the conditioner block ($M=m+1$). Encoding the embedded state as one letter over $\tilde\alpha=\alpha^{M}$ is the multivariate generalisation of Lempel-Ziv complexity to vector-valued processes \cite{zozor2005lempel}, and placing the target's present alongside the joint past makes (\ref{eq:embed_h}) a conditional rather than a joint rate \cite{res20}. Unlike a plain concatenation of the two strings, the embedding preserves the temporal alignment between the conditioner's past and the target's present, so the estimate responds to a lagged influence $w_j\to w_i$.

\section{Lempel-Ziv algorithm\label{sec:algol}}

We developed \textit{lzcomplexity} \cite{lzcomplexity}, an open-source library implemented in C++ and Python that computes the entropic magnitudes described from the Lempel-Ziv factorisation. The algorithms have been optimised to fully exploit current computing power by implementing parallel processing that significantly accelerates performance. The parallel implementation of the algorithms uses the Intel oneTBB library \cite{intelcorporationIntelOneAPIThreading2023}.

The algorithm used to obtain the factorisation consists of three phases \cite{crochemoreSimpleAlgorithmComputing2008}. The initial phase lexicographically sorts all suffixes of the input sequence. From the array of all sorted suffixes (SA), the lengths of the longest common prefixes (LCP) between consecutive suffixes are obtained. With the sorted suffixes, the second phase proceeds, where the longest previous factor (LPF) \cite{crochemoreComputingLongestPrevious2008} is computed for each position. Finally, the LZ76 factorisation is obtained from the LPF array. Details will be given elsewhere.

The implementation shows execution-time improvements of more than $10$-fold compared to sequential algorithms for long ($>10^6$) strings, and supports processing strings as large as $10^9$ symbols. The developed software can be obtained at \url{https://github.com/pleros-ai/lzcomplexity} \cite{lzcomplexity}.

\section{Case studies\label{sec:case}}

\subsection{Brain connectivity}

The human brain’s function relies on a complex interplay of organisation across multiple scales, from individual neurons to large-scale networks. The “connectome,” or the map of structural and functional connections between brain regions, is central to understanding cognition and behaviour. Abnormalities in this network are linked to various neurological and psychiatric conditions, making its study crucial for diagnosis and treatment. Inferring these connections from imaging data such as fMRI is challenging: current methods often rely on strong assumptions, are limited to linear dynamics, or are sensitive to user-defined parameters, obscuring the brain's inherently non-linear behaviour.

To overcome these limitations, we proposed a novel, model-free analytical approach grounded in entropic magnitudes introduced above. Instead of assuming linear correlations, the method characterises each region by its unpredictability (entropy density $\widehat h$) and its memory (effective measure complexity $\widehat E$), and uses the informational distance $\widehat d$ to assess functional connections. It thus detects both linear and non-linear dynamics without prior assumptions about the data, providing a generalisable characterisation of cognitive states across experimental conditions and signal types.

The fMRI data used in this study were sourced from the Human Connectome Project (HCP), for subjects performing a motor task cued for finger tapping, toe squeezing, or tongue movement. The brain was parcellated into $360$ regions of interest (ROIs) using the Glasser atlas~\cite{glasserMinimalPreprocessingPipelines2013} ($180$ ROIs per hemisphere). 

The data is discretised before estimating the entropic magnitudes. For each time series, the mean value is computed and used as the binarisation threshold. Binarisation results in the loss of information from the original signal, but it is assumed that the binarised strings still contain sufficient relevant information about the system's dynamics. If this is the case, it can only be asserted by the results themselves.

\begin{figure}[!t]
\centering
\includegraphics*[width=\textwidth]{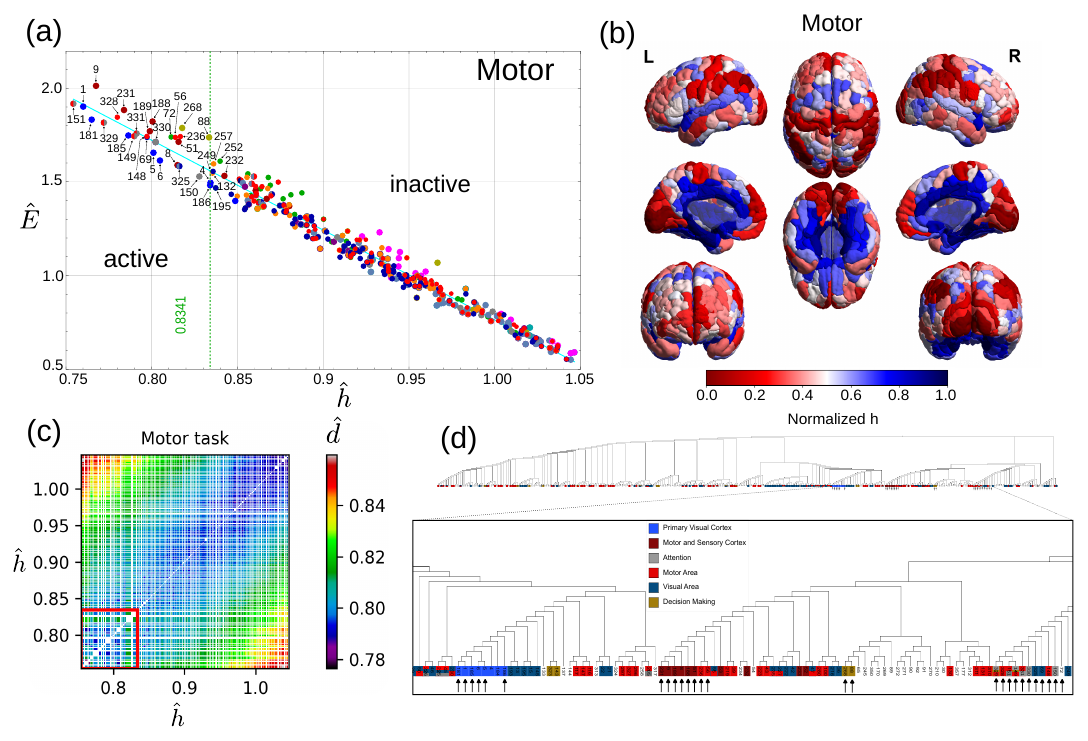}
\caption{\textbf{Brain connectome through the entropic magnitudes.}(a) The complexity-entropy map, $\widehat{E}$ vs $\widehat{h}$, for the $360$ brain parcellations during the motor task. Active regions (left of the vertical green line) cluster at lower $\widehat h$ and higher $\widehat E$, reflecting less unpredictability and stronger pattern formation. Inactive regions exhibit higher randomness, as they do not suppress the noisy background ~\cite{itoTaskevokedActivityQuenches2020}. (b) Brain surface mapping of entropy density showing the normalized $\widehat{h}$ for each ROI projected onto the cortical surface, red indicates lower $\widehat{h}$ (more active); blue indicates higher $\widehat{h}$(less active). The visual regions, motor and sensory cortex are activated during the task. (c) The distance map. The informational distance was computed between all region pairs via equation (\ref{eq:dclz}) and ordered by activation level. Active regions (red square) lie farthest from non-active ones, while non-active regions cluster together. (d) Dendrogram from the distance matrix illustrating hierarchical relationships. Arrows point to active regions, indicating clusters of visual, motor, and attentional processing, with strong associations among them.}\label{fig:brain}
\end{figure}

Figure~\ref{fig:brain}(a) shows the complexity-entropy of the different regions of the brain in the motor task. For this task all regions lie above $\widehat h = 0.75$ and $\widehat E = 0.5$, placing them in a regime between complete predictability and total randomness, the hallmark of complex dynamics. The points form a continuum rather than distinct clusters, so no fixed cut separates active from inactive regions. We segment the map objectively from the residuals of a linear fit: a local threshold is computed for each point's neighbourhood, cleanly splitting the map in two. Informed by previous studies suggesting that inactive regions exhibit more random signal patterns~\cite{itoTaskevokedActivityQuenches2020}, the side of the map with lower entropy density (and higher complexity) is taken as active. This criterion identifies a small subset of regions for every task; for the motor task, $8.61\%$ of regions are classified as active.

The position of the active regions relative to the linear fit carries functional information. Most primary and sensory motor regions lie above the fit, generating more pattern complexity than the trend predicts for their entropy density, whereas the main visual areas lie below it. Reading the regression as the mean $\widehat E$--$\widehat h$ relation, the motor and sensory cortices ``overperform'' in pattern generation while the visual areas ``underperform'' during this task: the motor response demands precisely coordinated movement, whereas the sustained visual cue allows looser visual engagement. This separation of regions that share an activation status but differ in dynamical regime is information that a single-amplitude activation measure cannot provide.

Projecting the normalised $\widehat h$ onto the cortical surface (Figure~\ref{fig:brain}(b)) shows that the active set is not scattered but anatomically coherent. The lowest $\widehat h$ values concentrate in the occipital visual areas and in the motor and sensory cortices of the posterior precentral gyrus, the two systems the task engages: a visual cue followed by a coordinated movement of fingers, toes, or tongue. The activation pattern recovered by the entropic criterion matches the known substrate of the task, obtained without a design matrix, a hemodynamic model, or any tuned parameter.

Functional connections were obtained from the LZ-distance $\widehat d$ between all $360$ ROIs. The resulting distance matrix, ordered by increasing $\widehat h$ (Figure~\ref{fig:brain}(c)), was clustered with the Neighbour-Joining method to produce the dendrogram of Figure~\ref{fig:brain}(d). Regions with similar $\widehat h$ are close in $\widehat d$ (the diagonal band), but this band narrows sharply over the active block, indicating tighter functional coupling among active regions. Active regions cluster together at low levels of the dendrogram, and regions of similar functional roles group consistently: visual areas of the occipital lobe form one cluster, primary motor/sensory and premotor regions another, and a third, heterogeneous cluster integrates occipital, lateral-frontal, and lateral-parietal regions with mixed visual, motor, and attentional roles. The full region-by-region assignment is reported in the companion study~\cite{mesa2025entropy}.

This example uses the motor task to illustrate the pipeline. Applied unchanged to memory, emotion, language, and resting-state data, the same measures recover task-relevant activation and modular functional structure, and the active sets agree with GLM and meta-analytic references; these results are detailed in~\cite{mesa2025entropy}.

\subsection{Non-linear locally coupled oscillators}

Coupled non-linear oscillators under weak coupling have been extensively studied since the foundational work of Winfree \cite{winfre67}, which established the oscillator phase as the key dynamical variable, superseding amplitude-based descriptions. This framework was anticipated by Adler’s analysis of locking in feedback circuits \cite{adler73}, leading to the phase equation for Adler-type oscillators:
\begin{equation}
\frac{\mathrm{d}\theta}{\mathrm{d}t}=-\gamma \sin \theta +\Delta \omega, \label{eq:adler}
\end{equation}
where $\theta$ denotes the phase, $\gamma$ a control parameter, and $\Delta \omega$ the frequency detuning. This formulation was later generalised to ensembles of coupled oscillators, culminating in the Kuramoto model \cite{kuramoto75}, which describes globally coupled oscillators via harmonic signals and has become a paradigm for studying synchronisation across physics, biology, and neuroscience \cite{acebron05}.

\begin{figure}[!ht]
\centering
\includegraphics*[scale=0.6]{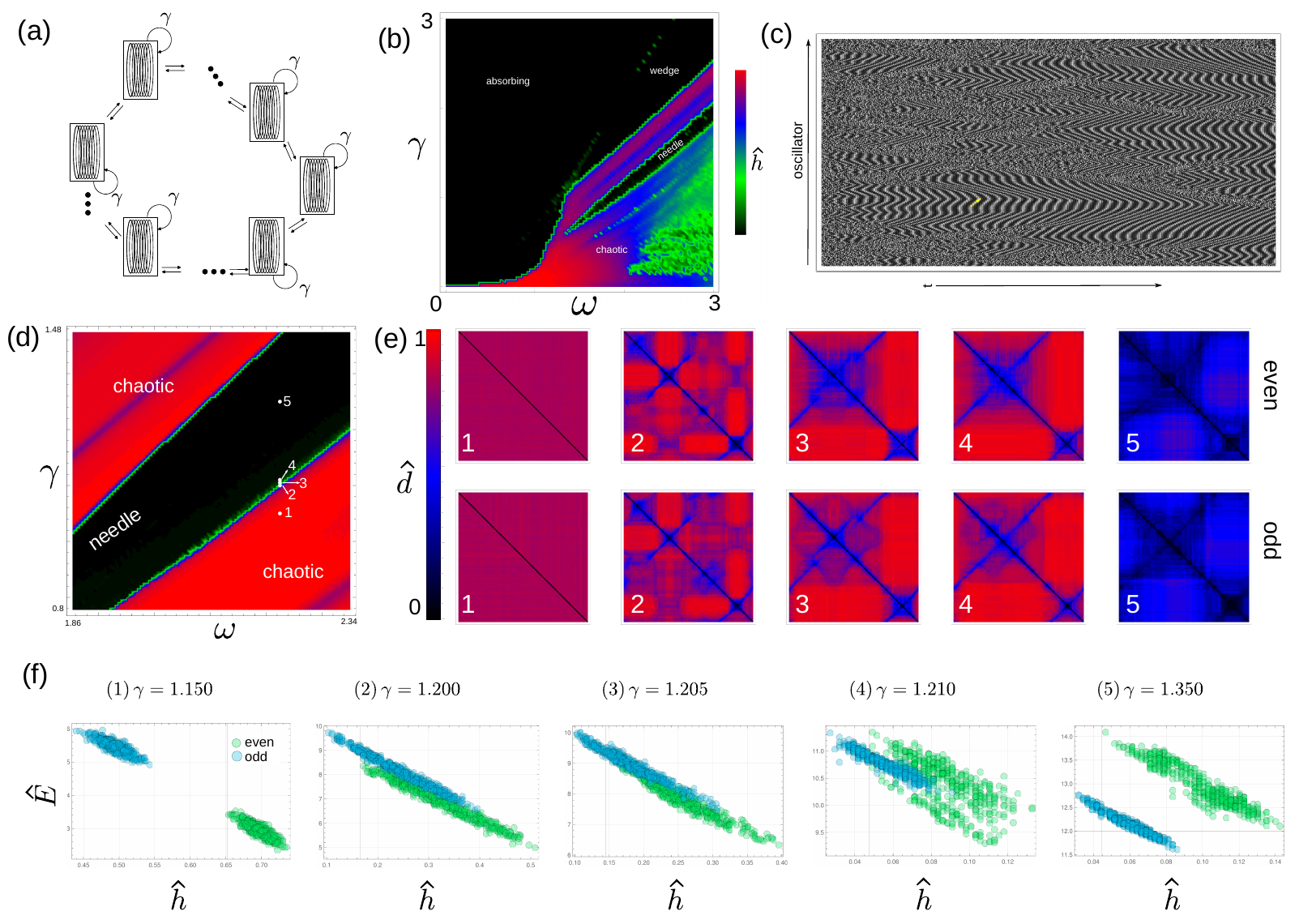}
\caption{\textbf{Non-linear locally coupled oscillators.} (a) Adler-type oscillators are coupled to their nearest neighbours as described by equation (\ref{eq:mastereq}), forming a circular chain. (b) The entropy density map as a function of the control parameters $\omega$ and $\gamma$. Four regions are identified by the entropic magnitudes. The absorbing region has zero entropy density, while the chaotic region has a value near $1$. For the wedge region, the entropy density is higher than the absorbing region but still lower than $1/2$. Low values of $h$ characterise the needle region. (c) Spatio-temporal map for $\gamma=1.206$ and $\omega=2.225$ at the border between the needle and the chaotic region (edge of chaos boundary). Wave-like patterns can be identified. Time is on the horizontal scale, and the oscillator’s values are displayed vertically in consecutive order. (d) Detail of the spatio-temporal map. Point $1$ is in the chaotic region, while point 5 is well within the needle region; all other points are at the border of the needle-chaotic regions. (e) The informational distance map for all points shown in (d) is split into even (above) and odd (below) oscillators. In the chaotic region (point 1), no distinguishable pattern is observed; by contrast, in the rest of the analysed points, characteristic patterns are seen in the maps. (f) The complexity-entropy map for each point shown in (d) in the chaotic region (point 1) shows that the odd and even oscillators are clearly segmented, with the even oscillators having higher $\widehat{h}$ values and lower $\widehat{E}$ values. The segmentation becomes less clear for points on the edge of the chaos boundary, but there is still a clear grouping for each type of oscillator. At point 5, well within the needle region, the two sets of oscillators are again clearly segmented, with higher $\widehat{E}$ values for both sets and smaller $\widehat{h}$ values.
}\label{fig:osc}
\end{figure}

Recent work has introduced a system of locally coupled Adler-type oscillators arranged in a ring with nearest-neighbour interactions as shown in Figure \ref{fig:osc}(a)\cite{alonso17}:
\begin{equation}
\frac{\mathrm{d}\theta_i}{\mathrm{d}t}=\omega+\gamma \cos \theta_i+(-1)^ i \left [ \cos \theta_{i-1}+\cos \theta_{i+1}\right ].\label{eq:mastereq}
\end{equation}
Equations (\ref{eq:mastereq}) are integrated numerically with a fourth-order Runge-Kutta method, and the entropic magnitudes are computed from the discretised phase activity $\sin\theta_i$, following Ref.~\cite{estevez18}. The phase activity is binarised with its mean value as threshold: values greater than or equal to the mean are mapped to $1$ and the remaining ones to $0$. The temporal evolution of each oscillator thus becomes a binary string, and the spatial configuration of the ring at a given time a binary string of length equal to the number of oscillators.

This model exhibits a diverse configuration diagram as seen by the $\widehat{h}$ value, as a function of two control parameters (Figure \ref{fig:osc}(b)), revealing regimes with distinct spatiotemporal dynamics: an absorbing region with near-zero entropy, a chaotic region with high sensitivity and disorder, a wedge region characterized by long-range correlations, and a needle region displaying intermediate complexity and intricate patterns \cite{estevez18}. Notably, the system exhibits a transition to chaos as the needle regions is abandoned while $\gamma$ decreases, where computational capabilities—namely information storage, transmission, and modification—are enhanced near the edge of chaos \cite{estevez19,estevez23,estevez24}, supporting the hypothesis that dynamical systems poised at this transition can function as optimized computational substrates \cite{packard88,langton90,kauffman92,bertschinger04}.

The needle region has an estimated entropy density near zero, while the effective measure complexity approaches $3$. At the lower boundary of the needle region, there is a surge in the $\widehat{E}$ value and a sudden fall to zero. It is at this surge point that enhanced system performance is expected. One such point is at $\omega=2.225$, $\gamma=1.206$, and the spatial-temporal diagram shown in Figure \ref{fig:osc}(c) exhibits a complex pattern with local spatial arrangements seen as wave-like patterns in the oscillator's behaviour, signalling that lasts for long temporal intervals.

When carefully studying the system behaviour with $\omega=2.225$ fixed and $\gamma$ varied (Figure \ref{fig:osc}(d)), several distinct pictures emerge. What is striking is that couplings seem to emerge between odd-indexed oscillators and between even-indexed oscillators, as seen in the $\widehat{d}$ maps shown in Figure \ref{fig:osc}(e). While at point 1, within the chaotic region no recognizable pattern is seen in the informational distance map, but the $\widehat{E}$ vs $\widehat{h}$ complexity-entropy map (Figure \ref{fig:osc}(f)) clearly shows that both set of oscillators, even and odd, can be segmented by their values of both entropic magnitudes. The even oscillators exhibit higher entropy density and lower $\widehat{E}$ values. As the studied points approach the edge and the needle region, clear patterns emerge in the distance map (points 2, 3, 4 in Figure \ref{fig:osc}(d)) for both sets of oscillators. The entropic behaviour of the two groups of oscillators converges, as evidenced by the entropy-complexity map (Figure \ref{fig:osc}(f)). Patterns in the distance map are a sign of coupling in the oscillators and therefore of collective behaviour. Well within the needle region, at point 5 in Figure \ref{fig:osc}(d) the entropy-complexity map again shows a clear segmentation of odd and even oscillators but, compared to what was seen in point 1, the range of $\widehat{h}$ values for both groups overlaps, and the difference is mostly seen as a gap between the $\widehat{E}$ values of both sets.

\subsection{Electrocardiogram analysis}

\begin{figure}[!ht]
\centering
\includegraphics*[width=\textwidth]{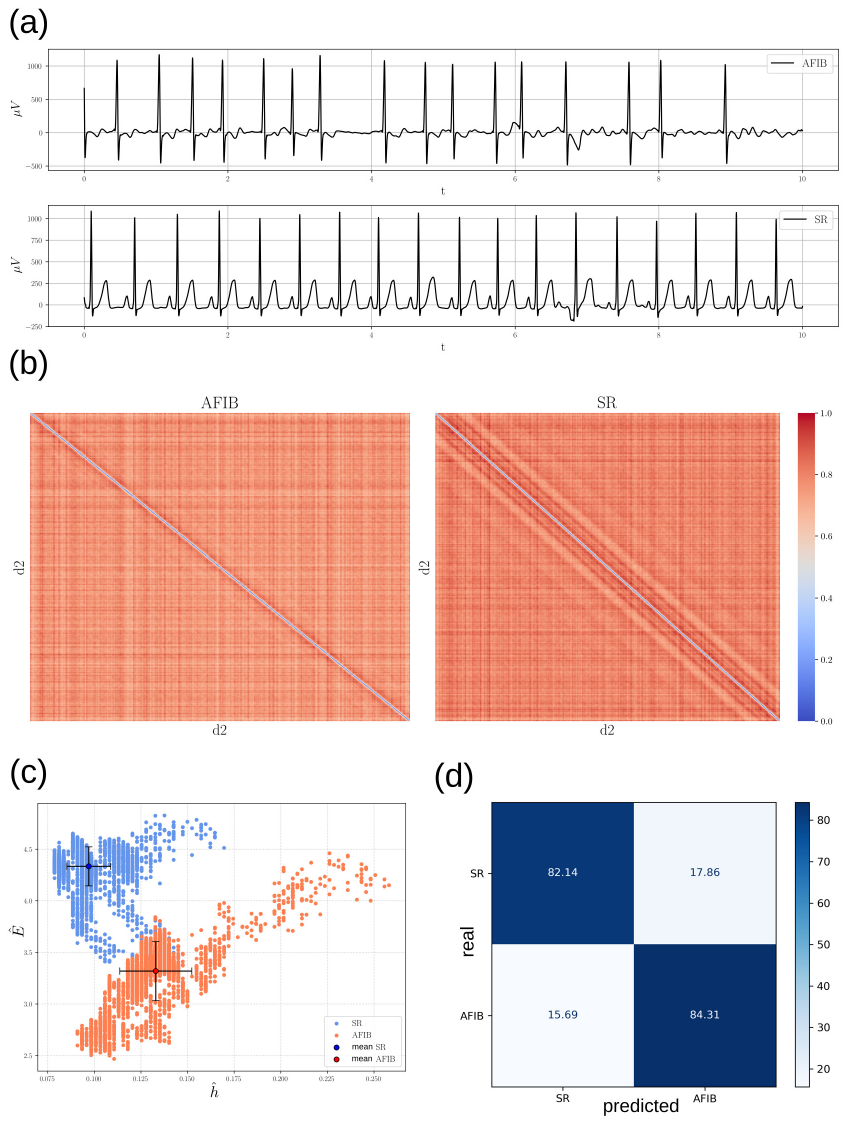}
\caption{\textbf{Electrocardiogram analysis.}(a) ECG of a healthy subject (below) and a subject with atrial fibrillation (above). The AFIB patient has irregular heart rate behaviour. (b) The recurrence map of the (left) AFIB patient and the (right) healthy subject. The SR subject's map shows clear, long diagonal lines parallel to the main diagonal, which are absent in the AFIB subject's map. (c) The complexity-entropy map, derived from the estimation of the entropy density $\widehat{h}$ and the effective measure complexity $\widehat{E}$ for each row of the recurrence matrix. SR and AFIB subjects are clearly segmented, with healthy subjects exhibiting higher $\widehat{E}$ and lower $\widehat{h}$. (d) Confusion matrix resulting from the training of a convolutional neural network fed with the vector of entropy density values. True negatives and true positives are above $82\%$.
}\label{fig:ecg}
\end{figure}

Electrocardiogram (ECG) classification has become a fundamental tool in modern cardiology. Usually, a complete ECG comprises signals from $12$ sensors or leads. However, manual interpretation of ECGs faces several challenges, including limited access to specialists in rural areas, inter-observer variability, and low sensitivity in short recordings. ECG is the most widely used tool for diagnosing cardiac diseases. Figure \ref{fig:ecg}(a) shows the ECG of a healthy subject (SR) and of a patient with atrial fibrillation (AFIB). There have been attempts to use machine learning techniques to diagnose such illnesses from ECGs. The most commonly used approach in this regard is to train the neural network directly on labelled ECG data and use the trained model for diagnosis. We followed a different approach.

From the ECG, the recurrence matrix is computed (Figure \ref{fig:ecg}(b)). An entry $r_{ij}$ in the recurrence matrix is the absolute value of the difference between the signal at time index $i$ and time index $j$:
\[
r_{ij}=|(ecg)_i-(ecg)_j|
\]

Figure \ref{fig:ecg}(b) shows the average recurrence matrices corresponding to AFIB and SR subjects.

Visually, the matrices reveal fundamental differences. The graph for the SR group exhibits a pattern of continuous diagonal lines parallel to the main diagonal. According to recurrence matrix theory, this characteristic indicates more deterministic, regular underlying dynamics, consistent with the periodic nature of SR. The presence of these diagonal lines indicates that trajectory segments evolve similarly at different time instants.

In contrast, the recurrence matrix obtained for the AFIB group suggests more complex cardiac dynamics, with a significant stochastic component. The lack of extended temporal recurrences in the trajectory reflects the inherent irregularity and loss of periodicity characteristic of this arrhythmia.

This confirms that diagonal lines constitute the key differentiating element between the two groups, serving as a visual biomarker of cardiovascular system dynamics. The abundance and continuity of these diagonal structures in the SR group not only capture the deterministic nature of SR but also quantify the system’s ability to maintain coherent and predictable temporal patterns.

The differences observed in the matrices suggest that entropic measures calculated from these structures will effectively capture their dynamic differences, thus providing a robust quantitative basis for automatic classification. From the distance matrices constructed for an AFIB subject and an SR subject, we proceed to calculate the entropic descriptors, entropy density and effective complexity measure for each row of the matrix. The values of $\widehat{E}$ and $\widehat{h}$ for each row for each lead are shown in the entropy-complexity map of Figure \ref{fig:ecg}(c)

In the healthy patient, values tend to cluster in a region of low $\widehat{h}$ and high $\widehat{E}$, as expected from our description of the dynamics of subjects with SR. We observe higher values of $\widehat{h}$ and lower values of $\widehat{E}$ for the AFIB subject, consistent with more random dynamics and reduced long-range correlations and system memory. The presence of a fraction of AFIB data that exhibits high entropy density $\widehat{h}$ values and, simultaneously, effective complexity measure $\widehat{E}$ values comparable to those of the SR subject is interesting; it could represent segments of the time series that regularise or show more patterns.

Following this analysis, a convolutional neural network (CNN) was trained using a feature vector as follows. The signals from two leads, lead II and lead V1, were used, as they are known to be significant for diagnosing AFIB. Three distance maps were computed: one for each lead and one for the cross-signal between the leads. From each recurrence map, the entropy density $\widehat{h}$ was estimated for each row from the binarised signal (binarisation was performed by the median). The entropy density for each row and recurrence map was treated as a vector, which was used to train the CNN. The number of control and AFIB subjects was roughly equal during training. Details of the CNN architecture are given in Appendix \ref{app:cnn}.

The network was trained using 5-fold cross-validation, with the data split into training and validation sets. The result is shown in Figure \ref{fig:ecg}(d) as a confusion matrix. The classification results obtained with this neural network provide strong evidence of the discriminative capacity of the extracted features (the vector $\widehat{h}$). In particular, this type of descriptor has proven highly representative of the underlying dynamics in ECG recordings, capturing complex aspects of nonlinear physiological behaviour that are traditionally inaccessible through conventional linear metrics. The performance achieved for $\widehat{h}$—with an overall accuracy of 83.2\% using only two leads of the 10-second ECG—is particularly relevant when considering that the most effective methods reported in the literature typically employ all 12 standard leads or long-duration ECGs with beat-to-beat annotations. This level of accuracy, achieved with a significantly reduced amount of information, highlights not only the efficacy of the proposed approach but also its practical viability in environments where access to multichannel data is limited or where more efficient signal acquisition and processing is required.

\subsection{Further applications\label{sec:other}}

We applied the same magnitudes to other systems in previous and ongoing work; those results are not reproduced here for lack of space but are summarised for completeness. In written language, $\widehat h$ and $\widehat E$ were estimated for the books of different writers, each text treated as a symbolic sequence of characters, without any grammatical or syntactic analysis. On the complexity-entropy map the authors occupy distinct regions, so the measures capture their different styles. Also, dendrograms built from the Lempel-Ziv distance matrix segmented the authors and each author's work by genre. Surrogates obtained by randomising each text at the sentence, word and character level allow the contribution of each level of organisation to be separated: the difference in $\widehat h$ between consecutive randomisations measures the information invested in organising characters into words, words into sentences, and sentences into the text, and the corresponding difference in $\widehat E$ measures the effective measure complexity gained at each of those levels, which decreases as one moves up the hierarchy \cite{estevez2019complexity}. 
 
In gene expression data, the magnitudes were estimated from whole-genome expression levels ordered by pathway, for thirteen tumour types taken from the TCGA repository. The complexity-entropy map separates tumour from normal samples in nine of the thirteen types, with success rates above $98\%$ in several of them, and $\widehat d$ discriminates in addition between tumour types, as shown for two kidney tumours. Neither gene preselection nor a model of the disease is required, and the ordering by pathway, rather than the identification of individual up- or down-regulated genes, is what makes the trend in the map emerge \cite{mesa2022cancer}. 
 
For trajectories, a two-dimensional curve is discretised with the Freeman chain code, which yields a sequence over an alphabet of eight symbols and requires no embedding or threshold. The magnitudes estimated from that sequence separate regular from chaotic orbits in the H\'enon-Heiles system, and, applied to centre-of-pressure trajectories from a public human balance dataset, provide the features with which a neural network classifies subjects by fall history under different postural conditions \cite{pena2025trajectory}. In all cases the procedure is the one followed above: discretise the signal, factorise it, and estimate $\widehat h$, $\widehat E$ and $\widehat d$ from the factorisation, without tuning parameters or assuming a model for the source.

\section{Conclusions}

In this article, we have revisited the fundamental concepts of entropy and related magnitudes for the analysis of sequential data, providing a unified treatment of their mathematical foundations within information theory and ergodic theory. We have established the relationships between block entropy, entropy density, entropy gain, effective measure complexity, and informational distance, clarifying their interpretations and convergence properties. The connection between Shannon's probabilistic entropy and Kolmogorov's algorithmic complexity has been emphasised, with Ziv’s theorem providing the crucial link that enables practical estimation via Lempel-Ziv factorisation.

The Lempel-Ziv factorisation has been examined in detail, demonstrating its utility for estimating entropic magnitudes from finite sequences. We have presented estimation procedures for entropy density, effective measure complexity via coverage-adjusted block entropies, informational distance, and the transfer entropy rate, including error estimates and convergence considerations. The accompanying open-source software library, lzcomplexity \cite{lzcomplexity}, implements parallel algorithms capable of processing sequences up to ($10^9$) symbols, making these analyses accessible for practical applications.

The three case studies illustrate the framework's broad applicability. In brain connectivity analysis, entropic measures revealed task-specific activation patterns and hierarchical functional organisation, providing a model-free approach to characterise cognitive states. The study of coupled nonlinear oscillators demonstrated that entropy density and effective measure complexity capture transitions between dynamical regimes and identify collective behaviour through patterns of informational distance. Finally, electrocardiogram analysis showed that entropy-based features extracted from recurrence matrices can achieve clinically relevant classification performance for atrial fibrillation with only two leads, highlighting the approach's practical diagnostic potential.

Taken together, these results demonstrate that the entropic magnitudes discussed—entropy density, effective measure complexity, and informational distance—provide a robust, theoretically grounded framework for characterising complex dynamics across diverse domains. Their estimation via Lempel-Ziv factorisation offers a practical, parameter-free methodology that captures both randomness and memory in sequential data, with promising applications in physics, biology, neuroscience, and medicine.

\section{Acknowledgments}

CITMA is acknowledged for financial support under the project CARDENT, grant PN223LH010-053. The MPI-PKS partially funded this work through its visitor program. EER thanks H. Kantz at MPI-PKS for the in-depth discussion and the excellent working environment.

\section{Data availability and  authors statement}

The data supporting the findings in the brain and heart case studies come from third parties and are publicly available from the Human Connectome Project (https://www.humanconnectome.org/) and the PhysioNet Database (https://physionet.org/). The \textit{lzcomplexity} software used to estimate the entropic magnitudes is openly available at \url{https://github.com/pleros-ai/lzcomplexity} and archived on Zenodo, \url{https://doi.org/10.5281/zenodo.23531093} \cite{lzcomplexity}.

\textbf{The authors have no conflicts to disclose.}

\appendix

\section{ Block entropy $H_\ell(u)$ is a non-decreasing function of $\ell$.\label{app:bentropy}}
\[
P(w_0^{\ell})=P(w_0^{\ell-1})P(w_{\ell}|w_0^{\ell-1})\leq P(w_0^{\ell-1}),
\]
then
\[
\begin{array}{ll}
H_{\ell+1}(u)&=-\sum\limits_{w^{(\ell+1)}\in u} P(w_0^\ell) \log P(w_0^\ell)\\
&=-\sum\limits_{w^{(\ell+1)}\in u} P(w_0^\ell)\log P(w_0^{\ell-1}) -\sum\limits_{w^{(\ell+1)}\in u} P(w_0^\ell) \log P(w_{\ell} | w_0^{\ell-1}).
\end{array}
\]
For the first term on the right-hand side, taking $w^{(\ell+1)}\in u$,
\[
\begin{array}{l}

-\sum\limits_{w_0\ldots w_{\ell}} P(w_0 \ldots w_{\ell}) \log P(w_0\ldots w_{\ell-1})=- \sum\limits_{w_0\ldots w_{\ell-1}}\left[ \log P(w_0\ldots w_{\ell-1}) \sum\limits_{w_{\ell}}P(w_0 \ldots w_{\ell}) \right]\\
= -\sum\limits_{w_0\ldots w_{\ell-1}} P(w_0\dots w_{\ell-1}) \log P(w_0\ldots w_{\ell-1})
= H_\ell(u).
\end{array}
\]
Where use has been made of $ \sum\limits_{w_{\ell}}P(w_0 \ldots w_{\ell})= P(w_0 \ldots w_{\ell-1})$.

Now,
\begin{equation}
 H_{\ell+1}(u)=H_\ell(u)-\sum\limits_{w^{(\ell+1)}\in u} P(w_0\ldots w_{\ell}) \log P(w_{\ell} | w_0,...,w_{\ell-1}).\label{eq:hhh}
\end{equation}
Taking into account that $\log P(\cdot)\leq 0$, each term on the last sum is positive and the non-decreasing behaviour of $H_\ell(u)$ follows.

\section{ Entropy gain converges faster than $h_\ell$.\label{app:convergence}}

It has already been shown that $\Delta H_\ell(u)$ is a non-increasing function of $\ell$, hence $\Delta H_\ell(u)\geq h$ for all $\ell$. From (\ref{eq:cadena}) will have
\begin{equation}
h_\ell(u)=\frac{1}{\ell}\sum\limits_{i=0}^{\ell-1}H(u_i|u^{(i)}).\label{eq:cadena1}
\end{equation}
Also
\[
H(u_i|u_{i-1}u_{i-2}\cdots u_1)\geq H(u_{\ell-1}|u_{\ell-2}u_{\ell-3}\cdots u_1)
\]
and
\[
\begin{array}{rl}
h_\ell&\geq  \frac{1}{\ell}\sum\limits_{i=0}^{\ell-1}H(u_{\ell-1}|u^{(\ell-1)})\\\\
&=\frac{1}{\ell}\ell H(u_{\ell-1}|u^{(\ell-1)})\\\\
&=\Delta H_\ell.
\end{array}
\]
from where
\begin{equation}
h_\ell\geq \Delta H_\ell\geq h. \label{eq:faster}
\end{equation}

Using equation (\ref{eq:entgain1})
\begin{equation}
\frac{H_{\ell+1}}{\ell+1}=\frac{H_\ell}{\ell}-\frac{1}{\ell+1}\left(\frac{H_\ell}{\ell}-H(u_\ell|u^{(\ell)})\right). \label{eq:dhlx}
\end{equation}
If we prove that the term in parentheses on the right-hand side is non-negative, then $h_\ell(u)$ is a non-increasing function of $\ell$. Considering
\[
\frac{H_\ell(u)}{\ell}\geq H(u_\ell|u^{(\ell)}).
\]
and taking into account, by the chain rule, that
\[
H_\ell(u)=H(u_0)+H(u_1|u_0)+H(u_2|u_0u_1)+\cdots H(u_{\ell-1}|u^{(\ell-1)})
\]
it follows that
\[
\frac{H_\ell(u)}{\ell}=\frac{1}{\ell}\sum\limits_{i=0}^{\ell-1}H(u_i|u^{(i)})\geq H(u_{\ell-1}|u^{(\ell-1)}),
\]
where the last inequality is a consequence of the arithmetic mean always being greater than or equal to the smallest summand. $\Delta H_\ell$ being non-increasing, $h_\ell\geq  \Delta H_\ell \geq \Delta H_{\ell+1}$.

\section{$\liminf\limits_{\ell\rightarrow \infty} \zeta_\ell(u)$ and $\limsup\limits_{\ell\rightarrow \infty} \zeta_\ell(u)$ exist and are finite\label{app:limsup}}

Let $\inf\limits_{\ell>\ell_i} \zeta_\ell(u)$ be the infimum of $ \zeta_\ell(u)$ for $\ell>\ell_i$,
\[
 \zeta_\ell(u)\ge \inf\limits_{\ell'>\ell_i}  \zeta_{\ell'}(u).
\]
It is clear that if $\ell_q>\ell_i$, then
\[
\inf\limits_{\ell>\ell_q}  \zeta_{\ell}(u)\ge \inf\limits_{\ell>\ell_i}  \zeta_{\ell}(u).
\]
Let $\zeta^{\inf}_n=\inf\limits_{\ell>\ell_n}  \zeta_{\ell}(u)$ and assume that $q>i$ implies $\ell_q>\ell_i$, then $\zeta^{\inf}_n$ is a non-decreasing sequence of $n$, but inequality (\ref{eq:hsigma}) implies that $\zeta^{\inf}_n\leq 1$. Therefore the $\lim\limits_{n\rightarrow\infty}\zeta^{\inf}_n$ exists and furthermore
\[
\liminf\limits_{\ell\rightarrow\infty}  \zeta_{\ell}(u)=\lim\limits_{n\rightarrow \infty}\zeta^{\inf}_n \leq 1.
\]

The same logic implies that $\limsup\limits_{\ell\rightarrow\infty}  \zeta_{\ell}(u)$ also exists. For $\lim\limits_{\ell\rightarrow\infty}  \zeta_\ell(u)$ to exist, it must hold that
\begin{equation}
\zeta_\infty(u)=\lim\limits_{\ell\rightarrow\infty} \zeta_\ell(u)=\liminf\limits_{\ell\rightarrow\infty}  \zeta_{\ell}(u)=\limsup\limits_{\ell\rightarrow\infty}  \zeta_{\ell}(u).\label{eq:hlim}
\end{equation}
Both limits indeed coincide. Every word of length $\ell+m$ occurring in $u$ is the concatenation of a word of length $\ell$ and a word of length $m$, both occurring in $u$, so that $\#_{\ell+m}\leq \#_\ell\,\#_m$. The sequence $a_\ell\equiv\log\#_\ell=\ell\,\zeta_\ell(u)$ is therefore subadditive, $a_{\ell+m}\leq a_\ell+a_m$, and by Fekete's lemma \cite{fekete1923} the limit of $a_\ell/\ell$ exists and equals $\inf_\ell a_\ell/\ell$:
\[
\lim\limits_{\ell\rightarrow\infty}\zeta_\ell(u)=\inf\limits_{\ell\geq 1}\zeta_\ell(u),
\]
which, by inequality (\ref{eq:hpapa}), lies in $[0,1]$. Equation (\ref{eq:hlim}) follows. Note that inequalities (\ref{eq:hl1}) and (\ref{eq:hl1b}) only imply $|\zeta_{\ell+1}-\zeta_\ell|\leq 1/(\ell+1)\rightarrow 0$, which by itself would not guarantee convergence.

\section{Ziv theorem 5 \label{app:ziv78}}

To prove (\ref{eq:ziv5}), consider the set of all subsequences $u^n_1$ of length $n$ emitted by the ergodic source. By the asymptotic equipartition property, we have
\[
\lim\limits_{n\rightarrow \infty}P\left[\left|-\frac{1}{n}\log P(u_1^n)-h\right|>\epsilon\right]=0,
\]
For any $\epsilon>0$, arbitrarily small. Therefore, there exists an integer $\ell$, such that for all $n\geq \ell$
\[
P\left[\left|-\frac{1}{n}\log P(u_1^n)-h\right|>\epsilon\right]\leq \epsilon.
\]
For $n=\ell$ there exists a set $\Omega$ of typical sequences that includes at most $\alpha^{\ell(h+\epsilon)}$ elements and at least $(1-\epsilon)\alpha^{\ell(h-\epsilon)}$ elements, almost all equiprobable, such that
\begin{equation}
P\left(u_1^\ell\in\Omega\right)\geq 1-\epsilon.\label{eq:pu}
\end{equation}
Consider the following partition of $u$
\[
 u=u_1^{k-1}u_{k}^{\ell+k-1}u_{\ell+k}^{2\ell+k-1}\ldots u_{j\ell+k}^{(j+1)\ell+k-1}\ldots \;\;\;\;1\leq k\leq \ell,\;\;j=0,1,\ldots.
\]
Let $\mathds{1}_k(0)=0$,
\[
 \mathds{1}_k(j)=\left\{\begin{array}{ll}0, & u_{j\ell+k}^{(j+1)\ell+k-1}\in \Omega\\\\ 1, & \text{otherwise}\end{array}\right.
\]
Define
\[
\rho_k(u)=\lim\limits_{L\rightarrow \infty}\frac{1}{L}\sum\limits_{j=0}^{L-1}\mathds{1}_k(j).
\]
which exists due to the stationary nature of the sequence $\mathds{1}_k(1),\ldots \mathds{1}_k(i),\ldots$. Noting, from (\ref{eq:pu}), that $P\left(u_1^\ell\notin\Omega\right)\leq \epsilon$ we will have with probability $1$, then,
\[
\frac{1}{\ell}\sum_{k=1}^\ell\rho_k(u)=1-P[\Omega]\leq \epsilon,
\]
the equality due to the source's ergodicity. Therefore, there exists a $k_0$ ($1\leq k_0\leq \ell$) for which
\begin{equation}
\rho_{k_0}(u)\leq \epsilon.\label{eq:dmu}
\end{equation}
Assume, without loss of generality, that $k_0=1$ and consider
\[
u=u_1^\ell u_{\ell+1}^{2\ell}u_{2\ell+1}^{3\ell}\ldots u_{j\ell+1}^{(j+1)\ell}\ldots \;\;\;\;j=0,1,\ldots
\]
Construct the sequence $w$ where each block $u_{j\ell+1}^{(j+1)\ell}$ of $u$ that does not belong to $\Omega$ is replaced by a block that does belong to $\Omega$. From (\ref{eq:dmu}), it is clear that the Hamming distance between $u$ and $w$, $d_H(u,w)\leq \epsilon$.

By construction, the number of distinct $\ell$-blocks in $w=w_1^\ell w_{\ell+1}^{2\ell}\ldots$ cannot be greater than $ \alpha^{\ell(h+\epsilon)}$, therefore
\[
 \alpha^{\ell q \zeta_{\ell q}(w)}\leq \left[ \alpha^{\ell(h+\epsilon)}\right]^{q-1}\alpha^\ell=\alpha^{q\ell(h+\epsilon)-\ell(h+\epsilon)+\ell}\leq \alpha^{q \ell\left[ h+\epsilon + 1/q \right ] },
\]
or
\[
\zeta_{\ell q}(w)\leq \left[h+\epsilon+1/q\right].
\]
Hence
\begin{equation}
\zeta_\infty(w)=\lim\limits_{q\rightarrow \infty}\zeta_{q\ell}(w)\leq h+\epsilon.
\end{equation}
Now choose the $w_{j\ell+1}^{(j+1)\ell}$ that replace the corresponding blocks in $u$, such that
\[
\zeta_\infty(u)=\zeta_\infty(w)\leq h+\epsilon,
\]
and therefore, for $\epsilon\rightarrow 0$
\begin{equation}
\zeta_\infty\leq h.\label{eq:hli1}
\end{equation}

For the reverse inequality, we use the other half of the asymptotic equipartition property $|\Omega|\geq (1-\epsilon)\alpha^{\ell(h-\epsilon)}$. Since our source is ergodic, every block of $\Omega$ occurs in $u$ with probability $1$, so that,
\[
\alpha^{\ell\zeta_\ell(u)}\geq (1-\epsilon) \alpha^{\ell(h-\epsilon)}.
\]
Hence
\[
\zeta_\ell(u)\geq h-\epsilon+\frac{\log (1-\epsilon)}{\ell},
\]
taking the limit $\ell\rightarrow \infty$ and then $\epsilon\rightarrow 0$ gives
\begin{equation}
\zeta_\infty\geq h\label{eq:hli2}
\end{equation}

From (\ref{eq:hli1}) and (\ref{eq:hli2}) it follows
\begin{equation}
\zeta_\infty=h, \label{eq:zetah}
\end{equation}
in probability, which is what we wanted to prove.

\section{The exhaustive history of a sequence is the smallest of all history factorisations \cite{lempelComplexityFiniteSequences1976}\label{app:ex}}
Consider a sequence $w$ out of an alphabet $\mathcal{A}$ and a non-exhaustive history of $w=f_1f_2\ldots f_{n_F'}$. Then, for each factor $f_i$, these two conditions are fulfilled
\begin{equation}
\begin{array}{c}
f_1\in \mathcal{A}\\\\
f_i\pi\in \mathcal{V}(f_1f_2\ldots f_i\pi^2).\label{eq:ne}
\end{array}
\end{equation}
Consider the LZ-factorisation (exhaustive history) $w=F_1F_2\ldots F_{n_F}$. For each factor $F_i$, these three conditions are true,
\begin{equation}
\begin{array}{c}
F_1\in \mathcal{A}\\
F_i\pi\in \mathcal{V}(F_1F_2\ldots F_i\pi^2)\\\\
F_i\notin \mathcal{V}(F_1F_2\ldots F_i\pi).\label{eq:e}
\end{array}
\end{equation}
For all $1\leq i \leq \min(n_F, n_F')$,
\begin{equation}
|f_1f_2\ldots f_i|\leq |F_1F_2\ldots F_i|.\label{eq:ineq}
\end{equation}
Indeed, if this is not the case, consider $i_0$ the smallest index where the inequality breaks $|f_1f_2\ldots f_{i_0}|>|F_1F_2\ldots F_{i_0}|$. It follows that $F_{i_0}$ can not be a suffix of $f_{i_0}$, because then $|f_1f_2\ldots f_{i_0}|=|F_1F_2\ldots F_{i_0}|$ which has been excluded. A non-null suffix of $F_{i_0}$ must belong to $f_{i_0}$, otherwise $i_0$ will not be the smallest $i_0$ where the inequality breaks, then $|f_{i_0}|\geq 2$.

Because of condition (\ref{eq:ne}) and that $F_{i_0}$ is not a suffix of $f_{i_0}$, the following is true,
\[
F_{i_0}\in \mathcal{V}(f_1f_2\ldots f_{i_0}\pi^2).
\]
Let us consider two cases, first that $f_1f_2\ldots f_{i_0}\pi=F_1F_2\ldots F_{i_0}$, then
\[
F_{i_0}\in \mathcal{V}(F_1F_2\ldots F_{i_0}\pi),
\]
which contradicts the third condition in (\ref{eq:e}) of the exhaustive nature of $F_{i_0}$.

The second case takes  $|f_1f_2\ldots f_{i_0}\pi|>|F_1F_2\ldots F_{i_0}|$ which implies $f_{i_0}\in \mathcal{V}(f_1f_2\ldots f_{i_0}\pi)$ and then $F_{i_0}$ is not the longest prefix of $F_{i_0}\ldots F_{n_F}$ which has not occurred before, which is a contradiction with the exhaustive nature of $F_{i_0}$. This proves condition (\ref{eq:ineq}), from where $n_F \leq n_F'$ follows.

\section{Upper bound for the length of the Lempel-Ziv factorisation \cite{lempelComplexityFiniteSequences1976}. \label{app:bound}}

The inequality to be proven is
\begin{equation}
C_{LZ}(w)<\frac{N}{(1-\epsilon_N)\log N},
\end{equation}
where
\begin{equation}
\epsilon_N=2\frac{1+\log\log (\alpha N)}{\log N}.
\end{equation}

Consider a sequence of length $N$ and let $n_F$ be the number of different words in which  the sequence can be parsed.  By design $C_{LZ}(w)\leq n_F+1$. Consider the string $L$ formed by the concatenation of words lexicographically ordered up to a word length $k$, (e.g. $0\cdot1\cdot00\cdot01\cdot10\cdot11\ldots \cdot 11111$, $k=5$).

The length $N_k$ of the string $L$ is given by
\begin{equation}
N=N_k=\sum\limits_{i=0}^{k}i \alpha^i=\frac{\alpha}{\alpha-1}\left [\frac{1}{\alpha-1}+\alpha^k\left(k-\frac{1}{\alpha-1}\right)\right], \label{eq:nk}
\end{equation}
it follows
\begin{equation}
\Delta^{\max}_{k}=N_{k+1}-N_{k}=(1+k)\alpha^{k+1}.\label{eq:deltakm}
\end{equation}
The number $n_F=n_{F,k}$ for such a string is
\[
n_{F,k}=\sum\limits_{i=1}^{k}\alpha^i=\frac{\alpha}{\alpha-1}\left(\alpha^k-1\right),
\]
and
\begin{equation}
C_{LZ}(w)< n_{F,k}+1=\frac{\alpha}{\alpha-1}\alpha^k-\frac{\alpha}{\alpha-1}<\frac{\alpha}{\alpha-1}\alpha^k.\label{eq:nk1}
\end{equation}
On the other hand, from (\ref{eq:nk}),
\[
N_k\frac{\alpha-1}{\alpha}=\frac{1}{\alpha-1}+\alpha^k\left(k-\frac{1}{\alpha-1}\right)>\alpha^k\left(k-\frac{1}{\alpha-1}\right),
\]
from where
\begin{equation}
\alpha^k<\frac{\alpha-1}{\alpha}\frac{N_k}{k-\frac{1}{\alpha-1}}. \label{eq:vs}
\end{equation}
It follows from expression (\ref{eq:nk1}),
\begin{equation}
C_{LZ}(w)<\frac{N_k}{k-\frac{1}{\alpha-1}}\leq\frac{N_k}{k-1}. \label{eq:clz1}
\end{equation}

Consider a positive number $N$, and $N_k$ be the largest number given by (\ref{eq:nk}) such that $N_k<N$, then it can be written
\[
N=N_k+\Delta_k,
\]
and by equation (\ref{eq:deltakm}),
\[
0\leq \Delta_k<\Delta_{k}^{\max}.
\]
The number of new distinct words that appear in the sequence after increasing the length by $\Delta_k$ can not be larger than $\Delta_k/(k+1)$; therefore, for any $n$, using inequality (\ref{eq:clz1})
\begin{equation}
C_{LZ}(w)<\frac{N_k}{k-1}+\frac{\Delta_k}{k+1}<\frac{N_k+\Delta_k}{k-1}=\frac{N}{k-1}. \label{eq:clz2}
\end{equation}

The following is also true,
\begin{equation}
\log N\geq\log \sum\limits_{i=1}^{k}i\alpha^i>\log\left[\alpha^k\left(k-\frac{1}{\alpha-1}\right)\right]=k+\log\left(k-\frac{1}{\alpha-1}\right)>k,\label{eq:logn1}
\end{equation}
where expression (\ref{eq:vs}) has been used. On the other hand
\begin{equation}
\log N<\log\sum\limits_{i=1}^{k+1}i\alpha^i<\log\alpha^{k+1}\sum\limits_{i=1}^{k+1}i=k+1+\log\frac{(k+1)(k+2)}{2}<k+1+2\log (k+1).\label{eq:logn2}
\end{equation}
From this last inequality
\begin{equation}
k-1\geq \log N-2-2\log(k+1)>\log N-2\left[1+\log(1+\log N)\right]=(1-\epsilon_N)\log N,\label{eq:k1}
\end{equation}
where the equation (\ref{eq:logn1}) was used.

From the last inequality (\ref{eq:k1}) and (\ref{eq:clz2}), follows what we set to prove.

\section{Upper bound for the length of the Lempel-Ziv factorisation and entropy density.\label{app:length}}

The inequality to be proven is
\begin{equation}
C_{LZ}(w)\leq N h/\log N.
\end{equation}

The asymptotic equipartition property of an ergodic sources defines the ‘‘typical’’ set as those with entropy equal to the entropy rate and cardinality $\alpha^{h N}$. For a large enough sequence $N\gg 1$, almost all strings are ‘‘typical’’. We can then write
\begin{equation}
N_k=\sum_{i=0}^k i (\alpha^h)^i=\frac{\alpha^h}{\alpha^h-1}\left [\frac{1}{\alpha^h-1}+\alpha^{hk}\left(k-\frac{1}{\alpha^h-1}\right)\right]. \label{eq:nk11}
\end{equation}
Performing the same operation with $n_{F,k}$,
\begin{equation}
n_{F,k}=\sum_{i=1}^k(\alpha^h)^i=\frac{\alpha^h}{\alpha^h-1}\left(\alpha^{hk}-1\right). \label{eq:nks}
\end{equation}
now
\begin{equation}
C_{LZ}(w)\leq n_{F,k}+1 < \frac{\alpha^h}{\alpha^h-1}\alpha^{hk}. \label{eq:clzh}
\end{equation}
The same argument that led to inequality (\ref{eq:logn1}) can be applied to show that it remains true.

On the other hand,
\[
\log N\leq \log \sum_{i=1}^{k+1} i (\alpha^h)^i<\log\alpha^{h (k+1)}\sum_{i=1}^{k+1} i<h (k +1)+2\log(k+1).
\]
From where, 
\begin{equation}
h(k-1)>\log N-2 h-2\log(k+1). \label{eq:hkm1}
\end{equation}
Is already known that $C_{LZ}(w)< N/(k-1)$, using equations (\ref{eq:hkm1}) and (\ref{eq:logn1}),,
\begin{equation}
C_{LZ}(w)<\frac{N h}{(1-\epsilon'_N)\log N}, \label{eq:clze1}
\end{equation}
where
\[
\epsilon'_N=\frac{2 h}{\log N}+\frac{2\log(\log N+1)}{\log N},
\]
which tends to zero as $N\rightarrow \infty$. The property (\ref{eq:clzhN}) follows.

\section{Block entropy from a coverage-adjusted frequency estimator}\label{app:block}

Given a sequence $u$ the surrogate-and-Lempel-Ziv step is replaced with a direct, coverage-adjusted estimate of $H_\ell(u)$ from the block frequency histogram of the sequence itself, following Chao and Shen \cite{chaoNonparametricEstimationShannon2003} which avoid the need of building 
surrogate sequence and no Lempel-Ziv factorisation of it is taken: entropy computed from a frequency histogram is manifestly insensitive to the order in which the blocks occur, so the shuffling step was never intrinsic to estimating $H_\ell(u)$ -- it was only ever necessary to make an order-\emph{sensitive} estimator (Lempel-Ziv complexity) behave as if the blocks were exchangeable. Working from the histogram
directly removes that intermediate step and its associated bias.

Two choices complete the construction. First, the histogram is built from \emph{sliding}, overlapping length-$\ell$ windows of $u$ rather than a
non-overlapping partition: $m_\ell = N-\ell+1$ windows rather than $N/\ell$ blocks.
The two constructions estimate the same quantity -- the ergodic average
over the source's length-$\ell$ block distribution is, by definition,
insensitive to which of the $\ell$ possible phases a given window happens to
start on -- but the non-overlapping partition realises only one such
phase per sequence, which can measure something other than that average
for sources with structure commensurate with $\ell$: e.g. a sequence with
exact period $\ell$ has every non-overlapping $\ell$-block identical, so a
non-overlapping estimate of $H_\ell(u)$ reads exactly $0$ regardless of the
source's real per-block entropy, purely as an artifact of the single
phase sampled. Sliding by $1$ removes this phase-alignment sensitivity
entirely, and incidentally supplies far more samples per scale, directly
easing the $\#_\ell$-vs-$m_\ell$ imbalance identified above.

Second, entropy is read off each histogram with the Good-Turing/Horvitz-Thompson
coverage correction of Chao and Shen, rather than the maximum-likelihood estimate. Writing $n_i$ for the count of the $i$-th
distinct length-$\ell$ block observed among the $m_\ell$ windows, the
Good-Turing coverage estimate is $\widehat C_\ell = 1-f_{1,\ell}/m_\ell$, with
$f_{1,\ell}$ the number of blocks observed exactly once; each block's
coverage-adjusted probability is $\widehat p_i = \widehat C_\ell\,(n_i/m_\ell)$, and

\begin{equation}
\widehat H_\ell^{\mathrm{CS}} \;=\; -\sum_i \frac{\widehat p_i \log \widehat p_i}
{1-(1-\widehat p_i)^{m_\ell}},
\qquad \widehat H_0 \equiv 0,
\label{eq:HlhatCS}
\end{equation}

consistent with $H_0(u)=0$; here $\widehat H_\ell$ denotes $\widehat H_\ell^{\mathrm{CS}}$. The denominator is a Horvitz-Thompson correction for the categories the sample of size $m_\ell$ is likely to have missed entirely; unlike a first-order correction such as Miller-Madow, it remains reliable well into the regime $\#_\ell\gtrsim m_\ell$. Numerical results estimate that it recovers the true entropy to within $3\%$.

\subsubsection{Reliability safeguards: monotone projection and adaptive truncation}

Because $\Delta H_\ell(u) - h \geq 0$ for every $\ell$ (Appendix~\ref{app:convergence}),
the exact partial sum $\sum_{\ell\leq n}(\Delta H_\ell - h)$ is non-negative and
non-decreasing in $n$ -- a direct consequence of Appendix~\ref{app:bentropy}'s
monotonicity of $H_\ell(u)$ combined with Appendix~\ref{app:convergence}'s
bound. The finite-sample estimate $\widehat E(n)$ of Eq.~\eqref{eq:EhatCS} is
not guaranteed to inherit either property: $\widehat h$ still carries its own
finite-$N$ bias (Sec.~\ref{sec:est}, Eq.~\eqref{eq:amigo7}), and while
$\widehat H_\ell^{\mathrm{CS}}$ is markedly less biased than a Lempel-Ziv-based
estimate at any single $\ell$ (see above), $\widehat h$'s residual bias is still
multiplied by a growing $\ell$ in Eq.~\eqref{eq:EhatCS}: at large enough $n$,
this can accumulate into a visibly non-monotone, or even net-negative,
partial sum, even where every individual $\widehat H_\ell$ is itself accurate.

We therefore project the raw sequence $\{\widehat E(1),\ldots,\widehat E(n)\}$
onto the cone of non-negative, non-decreasing sequences before reading
off the answer, via the pool-adjacent-violators algorithm
\cite{ayerEmpiricalDistributionFunction1955}: consecutive values that
violate monotonicity are replaced by their pooled average, repeatedly,
until the whole sequence is non-decreasing, after which any remaining
negative values are floored at zero. The reported summands
$\Delta\widehat H_\ell - \widehat h$ are the increments of this projection rather
than the raw differences, and $\widehat E$ itself is the projected value at
$\ell=n$.

$n$ is further chosen as the index at which the raw
(unprojected) partial sum peaks, rather than the largest $n$ the
reliability bound of Sec.~\ref{sec:choice-of-n} below would allow: past
that peak, $\widehat h$'s residual bias dominates whatever genuine structure
remains to be resolved, and the decline that follows carries no
information the projection above should be allowed to fold back in. This
is a bound on what the data can support, not a claim of having recovered
$E$ in full: for sources whose correlation length is long relative to
what $N$ can resolve at the block sizes $\widehat H_\ell^{\mathrm{CS}}$ remains
reliable at, the true peak of the exact partial sum lies beyond where the
estimated one is truncated, and only a longer sequence recovers the difference. 

\section{Finite state automata\label{app:fsa}}

A finite-state process is one that can be represented by a finite-state automaton (FSA). In a finite-state automaton of the type considered here, a finite number of states ${S}$ are connected by transitions from one state to another. The transition from one state to another happens with a certain probability, and upon such a transition, a unique symbol is emitted. For a given state $S_i$, $\mathcal{T}^{a}_{ki}$ represents the probability of transitioning from state $i$ to state $k$ while emitting a symbol $a$. We will be interested in the class of processes where, given a state $S_i$, the emitted symbol $a$ determines the state to which the transition happens (unifilar), and we can write $\mathcal{T}^{a}_{i}=\mathcal{T}^{a}_{ki}$. The process starts in an initial state, and at each step, a transition to another state occurs while a symbol is emitted. At step $t$, the probability that the process is at state $S_i$ is denoted by $p_i^{(t)}$. The asymptotic probability of a state is denoted by $p_i^{(\infty)}$.

If one considers, among all possible FSAs, only the unifilar and minimal Markov chains \cite{zambella88}, then,
\[
h= -\sum\limits_i\sum\limits_a p_i^{(\infty)}\mathcal{T}^{a}_{i}\log \mathcal{T}^{a}_{i},
\]
and
\[
E=\sum\limits_{t=0}^{\infty}\sum\limits_{i}\sum\limits_{a}(p_i^{(\infty)}-p_i^{(t)})\mathcal{T}^{a}_{i}\log \mathcal{T}^{a}_{i}
\]

We have used three Markovian processes as examples: The even process, the 2-state symmetric process and the random-random-xor process.

\begin{figure}[!ht]
\centering
\includegraphics*[scale=0.8]{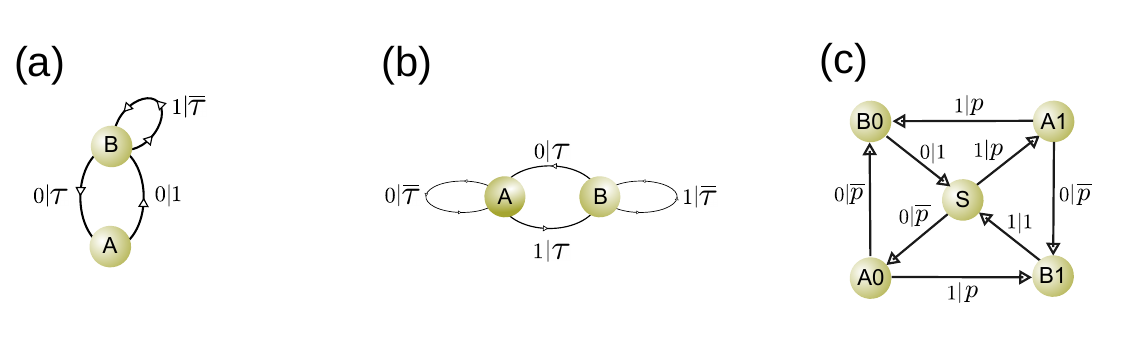}
\caption{Diagram of the recurrent part finite state automaton for the (a) even process; (b) two-state symmetric process and (c) the rrxor process.
}\label{fig:fsa}
\end{figure}

\textbf{The even process} can be represented by a finite state automaton as depicted in Figure \ref{fig:fsa}(a). It generates binary strings with an even number of $0$; the occurrence of a pair of $0$ happens with probability $\tau$, while the occurrence of the symbol $1$ happens with probability $\bar{\tau}=1-\tau$.

\textbf{The two state symmetric process (2ssp)} is shown in Figure \ref{fig:fsa}(b). It is a two-state FSA where the probability of transitioning from one state to the other is given by $\tau$, and, correspondingly, the probability of staying in the same state is $\bar{\tau}=1-\tau$.

\textbf{The random-random-XOR process (RRXOR)} is a five-state FSA, shown in Figure \ref{fig:fsa}(c). It generates binary strings in non-overlapping blocks of three symbols: the first two are independent, each equal to 1 with probability $p$ and to 0 with probability $\bar p = 1-p$; the third is fixed by the first two through $s_3 = s_1 \oplus s_2$.

\section{Convolutional neural network for heart disease diagnosis.\label{app:cnn}}

The structure of the CNN is given by:

\begin{itemize}
\item A first convolutional layer, with 64 filters and a kernel size of 5, designed to identify simple, low-level local patterns.
\item A ReLU activation function to introduce nonlinearity, allowing the network to learn more complex relationships.
\item A Max-pooling layer with a kernel size of 2.
\item A second convolutional layer, with 128 filters and a kernel size of 3. By having more filters and a smaller kernel, it seeks to combine the simple patterns detected by the first layer into more complex features.
\item Then a ReLU function and a Max-pooling layer are applied again, and subsequently the output is flattened.
\item Three fully connected layers: the first reduces dimensionality to 128 neurons, the second to 64, and the final one to a single output neuron. Between these
layers, ReLU activation functions and Dropout layers with a rate of 0.3 are applied.
\item The final output neuron uses a Sigmoid activation function. This function compresses the output to the range 0 to 1, allowing it to be interpreted as the probability that the patient belongs to each class.
\end{itemize}

\end{document}